\documentclass[times,sort&compress,3p, 12pt]{elsarticle}
\usepackage[labelfont=bf]{caption}

\usepackage{amsmath,amsfonts,amssymb,amsthm,booktabs,color,epsfig,graphicx,hyperref,url}
\usepackage{multirow}
\usepackage{tikz}
\usepackage{calc}
\usepackage{diagbox}
\usepackage{bm}
\usepackage{graphicx}
\usepackage{adjustbox}
\usepackage{tabularx}

\def\bSig\mathbf{\Sigma}

 \newcommand{\txr}{\textcolor{black}}

\theoremstyle{plain} % Theorem-like structures provided by amsthm.sty
\newtheorem{theorem}{Theorem}

\theoremstyle{definition}

\newtheorem{remark}{Remark}

\allowdisplaybreaks
\begin{document}

\begin{frontmatter}

\title{A non-parametric U-statistic testing approach for multi-arm clinical trials with multivariate longitudinal data}

\author[1]{Dhrubajyoti Ghosh\corref{mycorrespondingauthor}}
\author[1]{Sheng Luo}

\address[1]{Department of Biostatistics \& Bioinformatics, Duke University}
% \address[2]{Duke University}

\cortext[mycorrespondingauthor]{Corresponding author. Email address: dg302@duke.edu \url{}}

\begin{abstract}
Randomized clinical trials (RCTs) often involve multiple longitudinal primary outcomes to comprehensively assess treatment efficacy. The Longitudinal Rank-Sum Test (LRST) \cite{xu2025novel}, a robust U-statistics-based, non-parametric, rank-based method, effectively controls Type I error and enhances statistical power by leveraging the temporal structure of the data without relying on distributional assumptions. However, the LRST is limited to two-arm comparisons. To address the need for comparing multiple doses against a control group in many RCTs, we extend the LRST to a multi-arm setting. This novel multi-arm LRST provides a flexible and powerful approach for evaluating treatment efficacy across multiple arms and outcomes, with a strong capability for detecting the most effective dose in multi-arm trials. Extensive simulations demonstrate that this method maintains excellent Type I error control while providing greater power compared to the two-arm LRST with multiplicity adjustments. Application to the Bapineuzumab (Bapi) 301 trial further validates the multi-arm LRST's practical utility and robustness, confirming its efficacy in complex clinical trial analyses.
\end{abstract}


\end{frontmatter}

\section{Introduction\label{sec:1}}

Randomized clinical trials (RCTs) for neurodegenerative diseases, such as Alzheimer’s Disease (AD), aim to comprehensively evaluate treatment efficacy by capturing dynamic changes over time across various domains, including cognitive, functional, and behavioral assessments. Analyzing multiple longitudinal primary outcomes in these trials presents significant statistical challenges due to the need to account for correlations between outcomes and time points, as well as inherent variability in participant responses. Traditional statistical methods often require multiplicity adjustments to control Type I error rates, which can reduce statistical power and complicate the analysis. Additionally, these methods may rely on strict distributional assumptions, limiting their ability to fully leverage the complex multivariate nature of longitudinal data.

In the context of evaluating multivariate longitudinal outcomes from AD RCTs, two primary methodological approaches are prevalent. The first is the individual test approach, where each longitudinal outcome is analyzed independently or jointly, followed by multiplicity adjustments to control for Type I error. Techniques used in this approach include parametric methods like generalized estimating equations (GEE) and linear mixed models (LMMs), as well as nonparametric methods such as functional mixed models (FMMs) \cite{Zhu2019Stat-Sinica} and various rank-based tests \cite{Brunner2017JRSSB, Zhuang2018SMMR, Umlauft2019JMVA, Rubarth2022SMMR, Rubarth2022Symmetry}. However, these methods often require multiplicity adjustments that can reduce statistical power and require larger sample sizes. Moreover, LMMs may be biased by non-normality due to skewness and outliers \cite{Lachos2011Biometrics}, while global testing procedures relying on strict distributional assumptions may not fully utilize the available multivariate longitudinal data. The second approach transforms longitudinal data into a cross-sectional format, focusing on changes from baseline to the final observation. This method employs global testing procedures, categorized into parametric and nonparametric methods. Parametric procedures \cite{tang1994uniformly,lu2013halfline} often rely on assumptions like multivariate normality, which may be restrictive. In contrast, nonparametric rank-based procedures \cite{o1984procedures,Huang2005Biometrics,liu2010rank,zhang2019cluster}, being distribution-free, are widely used in clinical research, but these rank-based tests are primarily applied to cross-sectional data and are inadequate for tracking temporal variations in treatment efficacy, which is essential for a comprehensive assessment of treatment effects over time.

The Longitudinal Rank-Sum Test (LRST) \cite{xu2025novel} is a robust, U-statistics method that is rank-based and non-parametric, offering significant advantages over existing state-of-the-art approaches. By ranking data similarly to the Wilcoxon rank-sum test, LRST ensures robustness against various data distributions commonly encountered in clinical research. This method enhances statistical power, effectively controls Type I error, and reduces the required sample size by enabling a global assessment of treatment efficacy across multiple endpoints and time points without requiring multiplicity adjustments. Additionally, LRST fully leverages longitudinal data, capturing dynamic changes across multiple endpoints to provide a comprehensive analysis of treatment effects. However, the current LRST is limited to comparing two arms (treatment vs. control or placebo). In many RCTs, multiple doses are tested against a single control arm to identify the most effective dose. This multi-arm design presents additional challenges, particularly in controlling Type I error rates when evaluating multiple doses. Existing methods often require stringent multiplicity adjustments that further reduce power and complicate the analysis.

To address these challenges, we extend the LRST to a multi-arm setting, developing a novel multi-arm LRST that offers robust and flexible solutions for evaluating treatment efficacy across multiple treatment arms and a common control. This extension fully utilizes the longitudinal data, capturing the comprehensive temporal dynamics of treatment effects. By applying the multi-arm LRST, researchers can assess treatment efficacy globally across all arms without requiring multiplicity adjustments, thereby improving the understanding of dose-response relationships and optimizing trial designs. Upon rejecting the null hypothesis of no treatment efficacy, individual LRSTs (comparing the control to each dose group) can be conducted to identify the most efficacious dose for further investigation.

The article is structured as follows. Section~\ref{sec:2} outlines the proposed multi-arm LRST for multi-arm trials, with Section ~\ref{subsec:20} providing a brief overview of the two-arm LRST \cite{xu2025novel}. Section~\ref{subsec:21} introduces the methodology for three-arm trials, and Section~\ref{subsec:22} extends this approach to four-arm trials, with a discussion on generalizing the method to accommodate more arms. Section~\ref{sec:sim} presents an extensive simulation study demonstrating that our method not only achieves higher power but also effectively controls Type I error, outperforming the Bonferroni-corrected two-arm LRST. Finally, Section~\ref{sec:realData} applies the multi-arm LRST to the Bapineuzumab (Bapi) 301 trial, with mathematical proofs provided in \ref{sec:proofs}.

\section{Methods\label{sec:2}}
In this section, we start with an overview of the LRST for two-arm trials as developed in \cite{xu2025novel}. We then introduce the methodology for three-arm trials, followed by its extension to four-arm trials. Additionally, we provide insights into generalizing the approach to trials with more than four arms. Each subsection covers the statistical procedures, hypothesis testing framework, and the derivation of relevant test statistics and asymptotic distributions.

\subsection{LRST for 2-arm trial \label{subsec:20}}

In AD RCTs with multiple longitudinal outcomes and two parallel arms (treatment vs. control or placebo), the data structure is often of the type ($x_{itk}, y_{jtk}, t$), where $x_{itk}$ is the change in outcome $k$ ($k=1, \ldots K$ for a total of $K$ outcomes) from baseline to time $t$ ($t=1, \ldots T$, where $T$ is the number of after-baseline visits) for subject $i$ ($i=1, \ldots n_x$) in the control group, and $y_{jtk}$ is defined similarly for subject $j$ ($j=1, \ldots n_y$) in the treatment group. Here, $n_x$ and $n_y$ denote the sample sizes of the control and treatment groups, respectively. \textcolor{black}{The control and treatment groups are assumed to be independent, with no specific distributional assumptions imposed on the observed data}. For simplicity, we assume larger outcome values are clinically favorable. The outcome vectors $(x_{1tk}, \ldots, x_{n_xtk})$ and $(y_{1tk}, \ldots, y_{n_ytk})$, representing the observations from the control and treatment groups, are combined and ranked, with higher values receiving higher ranks. The resulting mid-ranks, $R_{xitk}$ and $R_{yjtk}$, are then used for further analysis. Mid-ranks are defined as standard ranks when there are no ties, or the average ranks for tied observations. 

To construct the LRST procedure, we follow Brunner \textit{et al} \cite{Brunner02book} and define the relative treatment effect on outcome $k$ at time $t$ as $\theta_{tk}=P(X_{tk}<Y_{tk})-P(X_{tk}>Y_{tk})$, where $X_{tk}$ and $Y_{tk}$ are the random variables corresponding to their observed values $x_{tk}$ and $y_{tk}$, \txr{with $F_{tk}$ and $G_{tk}$ representing their respective marginal distributions}. \textcolor{black}{This definition relies on pairwise comparisons, which are independent of any covariance structure between the control and treatment groups, due to the rank-based nature of the method}. We then calculate the overall relative treatment effect across all outcomes at time $t$ as $\theta_{t} = \frac{1}{K} \sum_{k=1}^{K} \theta_{tk}$ and the overall treatment efficacy across all outcomes and time points as $\bar{\theta} = \frac{1}{T} \sum_{t=1}^{T} \theta_t$. \textcolor{black}{Our approach assumes a directional alternative hypothesis (e.g., $\bar{\theta} > 0$), aligning with typical clinical trial objectives where treatments are expected to improve or slow disease progression compared to control. This design maximizes power for detecting directional alternatives, which are highly relevant in efficacy testing. However, testing general inequalities (e.g., $\bar{\theta} \neq 0$) may also be of interest. The LRST framework can accommodate such non-directional hypotheses by employing an $\ell_2$-based test statistic, as discussed in the Discussion Section.}. A positive value of $\bar{\theta}$ suggests that the treatment improves or slows the progression of some outcomes, leading to the following hypothesis test: 

\begin{equation}
H_0: \bar{\theta}=0  \quad vs \quad H_1: \bar{\theta}>0 \vspace{-1mm}
\label{eq:hypothesis_base}
\end{equation}

To test this hypothesis, \cite{xu2025novel} proposed the LRST statistic
\begin{equation}
\txr{\mathcal{T}_{LRST}}=\frac{\bar{R}_{y\cdot\cdot\cdot}-\bar{R}_{x\cdot\cdot\cdot}}{\sqrt{\widehat{var}(\bar{R}_{y\cdot\cdot\cdot}-\bar{R}_{x\cdot\cdot\cdot})}}, \vspace{-3mm}
\label{eq:LRST}
\end{equation} 
where $\bar{R}_{y\cdot\cdot\cdot}=\frac{1}{n_yTK}\sum_{j=1}^{n_y}\sum_{k=1}^{K}\sum_{t=1}^{T}R_{yjtk}$, $\bar{R}_{x\cdot\cdot\cdot}=\frac{1}{n_xTK}\sum_{i=1}^{n_x}\sum_{k=1}^{K}\sum_{t=1}^{T}R_{xitk}$ are the average ranks across all outcomes and time points in the treatment and control groups, respectively. Larger values of the group rank difference $\bar{R}_{y\cdot\cdot\cdot}-\bar{R}_{x\cdot\cdot\cdot}$ suggest treatment efficacy compared to the control. The denominator, $\widehat{var}(\bar{R}_{y\cdot\cdot\cdot}-\bar{R}_{x\cdot\cdot\cdot})$ is a consistent estimator of the variance of the group rank difference. 

The LRST statistic is an advanced rank-sum-type test that enhances existing rank-based tests \cite{Brien1984Biometrics, Huang2005Biometrics, Liu2010JASA, Zhang2019Biometrics} by 
effectively utilizing all available longitudinal data. Rejection of the null hypothesis $H_0$ in \eqref{eq:hypothesis_base} using the LRST statistic defined in \eqref{eq:LRST} indicates that the treatment is more effective than the control, providing directional inference. Following the rejection, LMMs may be applied to each outcome to identify and quantify specific treatment effects. For further details on the LRST procedure and its theoretical properties, please refer to \cite{xu2025novel}.

\subsection{Methods for three-arm trial\label{subsec:21}}

This section outlines the methodology for comparing three groups in a clinical trial: a control group and two treatment groups, typically representing two different doses of the same treatment. Building on the notation from Section~\ref{subsec:20}, let $x_{itk}$ be the change in outcome $k$ from baseline to time $t$ for subject $i$ in the control group. Similarly, $y_{jtk}$ denotes the corresponding change for subject $j$ ($j=1, \ldots, n_y$) in the low-dose group and $z_{ltk}$ for subject $l$ ($l=1, \ldots, n_z$) in the high-dose group. The random variables $X_{tk}$, $Y_{tk}$, and $Z_{tk}$ correspond to these observed values, with $F_{tk}$, $G_{tk}$, and $H_{tk}$ representing their marginal distributions. The total sample size $N$ is given by $N=n_x+n_y+n_z$, with $N_1=n_x+n_y$ and $N_2=n_x+n_z$. Under large-sample asymptotics, the ratios of each group's sample size to the total sample size are assumed to converge to constants, denoted as $\lambda_x$, $\lambda_y$, and $\lambda_z$. Similarly, the ratios $n_x/n_y \rightarrow \lambda_{x,y}$, $n_x/n_z \rightarrow \lambda_{x,z}$, $n_x/N_1 \rightarrow \lambda_x^y$, and $n_x/N_2 \rightarrow \lambda_x^z$ represent the corresponding limiting values. For simplicity, we assume that higher values indicate more favorable clinical outcomes. 

The three-arm LRST procedure is developed using pairwise ranking. First, we combine the observations of the control and low-dose groups for each outcome and time point, rank them, and obtain their mid-ranks denoted by $R_{xitk}^{(x,y)}$ and $R_{yjtk}^{(x,y)}$, respectively. Here, the superscript $(x,y)$ indicates pairwise ranking of the control and low-dose groups. The relative effect of the low-dose treatment compared to the control for outcome $k$ is defined as $\theta_{tk}^{(x,y)}=P(X_{tk}<Y_{tk})-P(X_{tk}>Y_{tk})$. The overall low-dose treatment effect across all outcomes at time $t$ is denoted by $\theta_t^{(x,y)} = \frac{1}{K}\sum_{k=1}^K \theta_{tk}^{(x,y)}$, with the vector of treatment effects over time represented as $\pmb{\theta}^{(x,y)} = (\theta_1^{(x,y)}, \ldots, \theta_T^{(x,y)})^T$. Similarly, we obtain the mid-ranks for the control and high-dose groups, $R_{xitk}^{(x,z)}$ and $R_{zltk}^{(x,z)}$, and define the relative effect of the high-dose treatment as $\theta_{tk}^{(x,z)}$, with corresponding overall treatment effects $\theta_t^{(x,z)}$ and $\pmb{\theta}^{(x,z)} = (\theta_1^{(x,z)}, \ldots, \theta_T^{(x,z)})^T$.

The null and alternative hypotheses of the overall test are as follows: 
\begin{equation}
H_0: \bar{\theta}^{(x,y)}=\bar{\theta}^{(x,y)}=0,  \quad \text{vs} \quad H_1: \bar{\theta}^{(x,y)} > 0 \text{ or } \bar{\theta}^{(x,z)}>0,
\label{nullxyz}
\end{equation}
where $\bar{\theta}^{(x,y)}=\frac{1}{TK}\sum_{t=1}^{T}\sum_{k=1}^{K}\theta_{tk}^{(x,y)}$ and $\bar{\theta}^{(x,z)}=\frac{1}{TK}\sum_{t=1}^{T}\sum_{k=1}^{K}\theta_{tk}^{(x,z)}$ representing the treatment efficacy of the low-dose and high-dose groups, respectively, compared to the control group. Rejection of this null hypothesis indicates that at least one treatment dose is effective compared to the control. To test these hypotheses in \eqref{nullxyz}, we propose a three-arm LRST test statistic denoted by \txr{$\mathcal{T}_{LRST}^{(x,y,z)}$} as in \eqref{eq:LRST_xyz}. 
\begin{align}
\mathcal{T}_{LRST}^{(x,y,z)} & = T^{-1}\max\left\{ \frac{\bar{R}^{(x,y)}_{y\cdot\cdot\cdot}-\bar{R}^{(x,y)}_{x\cdot\cdot\cdot}}{\sqrt{\widehat{var}(\bar{R}^{(x,y)}_{y\cdot\cdot\cdot}-\bar{R}^{(x,y)}_{x\cdot\cdot\cdot})}}, \frac{\bar{R}^{(x,z)}_{z\cdot\cdot\cdot}-\bar{R}^{(x,z)}_{x\cdot\cdot\cdot}}{\sqrt{\widehat{var}(\bar{R}^{(x,z)}_{z\cdot\cdot\cdot}-\bar{R}^{(x,z)}_{x\cdot\cdot\cdot})}} \right\}, \vspace{-10mm}
\label{eq:LRST_xyz}
\end{align} 
where $\bar{R}^{(x,y)}_{y\cdot\cdot\cdot}=\frac{1}{n_yTK}\sum_{j=1}^{n_y}\sum_{k=1}^{K}\sum_{t=1}^{T}R^{(x,y)}_{yjtk}$ and $\bar{R}^{(x,y)}_{x\cdot\cdot\cdot}=\frac{1}{n_xTK}\sum_{i=1}^{n_x}\sum_{k=1}^{K}\sum_{t=1}^{T}R^{(x,y)}_{xitk}$ are the average rank across all outcomes and time points in the low-dose and control groups, respectively. Similarly, we define $\bar{R}^{(x,z)}_{z\cdot\cdot\cdot}$ and $\bar{R}^{(x,z)}_{x\cdot\cdot\cdot}$. This statistic is calculated as the maximum of two standard normal random variables, each representing the difference between the average ranks of the low-dose and control groups, and the high-dose and control groups, respectively. A significant \txr{$\mathcal{T}_{LRST}^{(x,y,z)}$} suggests the efficacy of at least one dose over the control. This rank-max-type test extends the existing literature \cite{Liu2010JASA,Li2014SII,Meng2021SMMR,Zhang2022Bioinformatics} to longitudinal data.

To derive the asymptotic distribution of the three-arm LRST test statistic, we define the rank difference vector as $\bm{R}^{(x,y,z)}_{\text{diff}}=(\bm{R}^{(x,y)}_{\text{diff}}, \bm{R}^{(x,z)}_{\text{diff}})^{'}$, where 
$\bm{R}^{(x,y)}_{\text{diff}}=(\bar{R}^{(x,y)}_{y\cdot 1\cdot}-\bar{R}^{(x,y)}_{x\cdot 1\cdot},\cdots, \bar{R}^{(x,y)}_{y\cdot T\cdot}-\bar{R}^{(x,y)}_{x\cdot T\cdot})^{'}$ and $\bm{R}^{(x,z)}_{\text{diff}}=(\bar{R}^{(x,z)}_{z\cdot 1\cdot}-\bar{R}^{(x,z)}_{x\cdot 1\cdot},\cdots, \bar{R}^{(x,z)}_{z\cdot T\cdot}-\bar{R}^{(x,z)}_{x\cdot T\cdot})^{'}$ represent the differences in the average ranks between the low-dose and control groups and the high-dose and control groups, respectively. The covariance between the marginal distributions is defined as follows: 
\begin{align*}
\label{eqx}
 c_{t_1k_1t_2k_2}^{(x,y)} &= Cov\left(G_{t_1k_1}(X_{t_1k_1}), G_{t_2k_2}(X_{t_2k_2}) \right),  \quad
 c_{t_1k_1t_2k_2}^{(x,z)} = Cov\left(H_{t_1k_1}(X_{t_1k_1}), H_{t_2k_2}(X_{t_2k_2}) \right), \\
 d_{t_1k_1t_2k_2}^{(x,y)} &= Cov\left(F_{t_1k_1}(Y_{t_1k_1}), F_{t_2k_2}(Y_{t_2k_2}) \right),
 \quad d_{t_1k_1t_2k_2}^{(x,z)} = Cov\left(F_{t_1k_1}(Z_{t_1k_1}), F_{t_2k_2}(Z_{t_2k_2}) \right), \\
 c_{t_1k_1t_2k_2}^{(x,y,z)} &= Cov\left(G_{t_1k_1}(X_{t_1k_1}), H_{t_2k_2}(Y_{t_2k_2}) \right), 
 \quad \alpha_{x,y,z} = \left(1 +  \frac{1}{\lambda_{x,y}}\right)(\lambda_x + \lambda_z) = \left(1 +  \frac{1}{\lambda_{x,z}}\right)(\lambda_x + \lambda_y) .
 \end{align*}

\begin{theorem} \label{thm1}
The rank difference vector $\bm{R}^{(x,y,z)}_{\textnormal{diff}}$ is a multivariate U-statistic, and $\frac{1}{\sqrt{N}}\bm{R}^{(x,y,z)}_{\textnormal{diff}}$ asymptotically follows a multivariate normal distribution with mean $\frac{1}{2} \left( (\lambda_x + \lambda_y)\sqrt{N} \pmb{\theta}^{(x,y)}, (\lambda_x + \lambda_z)\sqrt{N} \pmb{\theta}^{(x,z)}\right)^{\top}$ and covariance matrix $\Sigma_{2 T \times 2 T}$ as $min(n_x,n_y,n_z) \rightarrow \infty$. The convergence matrix can be expressed as:

\[
\Sigma = 
\begin{pmatrix}
\Sigma^{(x,y)} & \Sigma^{(x,y,z)} \\
& \Sigma^{(x,z)}
\end{pmatrix},
\]
where $\Sigma^{(x,y)}_{t_1,t_2}$, $\Sigma^{(x,z)}_{t_1,t_2}$, and $\Sigma^{(x,y,z)}_{t_1,t_2}$ are defined by: 
\begin{align*}
 \Sigma^{(x,y)}_{t_1,t_2} &= (\lambda_x + \lambda_y) \frac{1}{K^2}\sum_{k_1,k_2 = 1}^K \left(\left(1 + \frac{1}{\lambda_x^y}\right)c_{t_1k_1t_2k_2}^{(x,y)} +  (1 + \lambda_x^y)d_{t_1k_1t_2k_2}^{(x,y)} \right),  \\ 
 \Sigma^{(x,z)}_{t_1,t_2} &=  (\lambda_x + \lambda_z)\frac{1}{K^2}\sum_{k_1,k_2 = 1}^K \left(\left(1 + \frac{1}{\lambda_x^z}\right)c_{t_1k_1t_2k_2}^{(x,z)} +  (1 + \lambda_x^z)d_{t_1k_1t_2k_2}^{(x,z)} \right), \\ 
  \Sigma^{(x,y,z)}_{t_1,t_2} &= \frac{1}{K^2}\sum_{k_1,k_2=1}^K 
  \alpha_{x,y,z}c_{t_1k_1t_2k_2}^{(x,y,z)}.
\end{align*}
\end{theorem}

The three-arm LRST test statistic \txr{$\mathcal{T}_{LRST}^{(x,y,z)}$} in \eqref{eq:LRST_xyz} can be written as
\begin{equation}
\label{maxTest}
    \txr{\mathcal{T}_{LRST}^{(x,y,z)}} = T^{-1}max \left\{\frac{\bar{R}^{(x,y)}_{y\cdot\cdot\cdot}-\bar{R}^{(x,y)}_{x\cdot\cdot\cdot}}{\sqrt{N J^T \hat{\Sigma}^{(x,y)} J}}, \frac{\bar{R}^{(x,z)}_{z\cdot\cdot\cdot}-\bar{R}^{(x,z)}_{x\cdot\cdot\cdot}}{\sqrt{NJ^T\hat{\Sigma}^{(x,z)} J}}\right\},
\end{equation}
where $J$ is a vector of $1$'s. Using the asymptotic results in Theorem~\ref{thm1}, we can obtain the asymptotic density of the test statistics \txr{$\mathcal{T}_{LRST}^{(x,y,z)}$} in Theorem~\ref{thm2}, with the estimators $\hat{\Sigma}^{(x,y)}$ and $\hat{\Sigma}^{(x,z)}$ being derived using consistent estimators of $c_{t_1k_1t_2k_2}^{(x,y)}$, $c_{t_1k_1t_2k_2}^{(x,z)}$, $d_{t_1k_1t_2k_2}^{(x,y)}$ and $d_{t_1k_1t_2k_2}^{(x,z)}$, the exact expressions of which are given in Theorem~\ref{thm3}. The p-value of the test can be obtained using this distribution. If the null hypothesis \eqref{nullxyz} is rejected, we examine the two components in \eqref{maxTest} to identify the dose level with a higher value, indicating greater efficacy.  

\begin{theorem}
\label{thm2}
    Under the conditions of Theorem~\ref{thm1}, the asymptotic density of 
 \txr{$\mathcal{T}_{LRST}^{(x,y,z)}$} is given by:
    \begin{align*}
        f(u) = 2 \phi \left( u \right) \Phi\left( u \sqrt{\frac{ 1- \rho}{1 + \rho}} \right)  ,
    \end{align*}
    where:
    \begin{align*}
        \rho = \frac{J^T \Sigma^{(x,y,z)} J }{\sigma_1 \sigma_2}, \quad 
        \sigma_1^2 &= T^{-2}J^T \Sigma^{(x,y)} J, \quad
        \sigma_2^2 = T^{-2}J^T \Sigma^{(x,z)} J.
    \end{align*}

    $\phi(\cdot)$ and $\Phi(\cdot)$ are the density and distribution functions of standard normal distributions.
\end{theorem}

Next, we provide the moment estimates of the underlying covariance terms. Suppose:
\begin{align*}
    \hat{\theta}_{tk}^{(x,y)} &= \frac{1}{n_x n_y} \sum_{i=1}^{n_x}\sum_{j=1}^{n_y}[I(x_{itk} < y_{jtk}) - I(x_{itk} > y_{jtk})] \\
    &= \frac{1}{n_x n_y}  \sum_{i=1}^{n_x}\sum_{j=1}^{n_y}[1 - 2 I(x_{itk} > y_{jtk})] = 1 - \frac{2}{n_x n_y} \sum_{i=1}^{n_x} \sum_{j=1}^{n_y} I(x_{itk} > y_{jtk}),
\end{align*}
and similarly, $\hat{\theta}_{tk}^{(x,z)} = \frac{1}{n_x n_z}\sum_{i=1}^{n_x}\sum_{l=1}^{n_z}[1 -2 I(x_{itk} > z_{ltk})]$.

\begin{theorem}
\label{thm3}
Under assumptions of Theorem~\ref{thm1}, the estimates of the covariance terms are given by:

\allowdisplaybreaks{
\begin{align*}
    \hat{c}_{t_1k_1t_2k_2}^{(x,y)} &= \frac{1}{n_x} \sum_{i=1}^{n_x} \left\{ \left(\frac{1}{n_y} \sum_{j=1}^{n_y} I(y_{jt_1k_1} < x_{it_1k_1}) - \frac{1 - \hat{\theta}_{t_1k_1}^{(x,y)}}{2}\right) 
    \left(\frac{1}{n_y} \sum_{j=1}^{n_y} I(y_{jt_2k_2} < x_{it_2k_2}) - \frac{1 - \hat{\theta}_{t_2k_2}^{(x,y)}}{2}\right)  \right\}, \\
    \hat{c}_{t_1k_1t_2k_2}^{(x,z)} &= \frac{1}{n_x} \sum_{i=1}^{n_x} \left\{ \left(\frac{1}{n_z} \sum_{j=1}^{n_z} I(z_{jt_1k_1} < x_{it_1k_1}) - \frac{1 - \hat{\theta}_{t_1k_1}^{(x,z)}}{2}\right)
    \left(\frac{1}{n_z} \sum_{j=1}^{n_z} I(z_{jt_2k_2} < x_{it_2k_2}) - \frac{1 - \hat{\theta}_{t_2k_2}^{(x,z)}}{2}\right)  \right\}, \\
     \hat{d}_{t_1k_1t_2k_2}^{(x,y)} &= \frac{1}{n_y} \sum_{j=1}^{n_y} \left\{ \left(\frac{1}{n_x} \sum_{i=1}^{n_x} I(x_{it_1k_1} < y_{jt_1k_1}) - \frac{1 + \hat{\theta}_{t_1k_1}^{(x,y)}}{2}\right) 
    \left(\frac{1}{n_x} \sum_{i=1}^{n_x} I(x_{it_2k_2} < y_{jt_2k_2}) - \frac{1 + \hat{\theta}_{t_2k_2}^{(x,y)}}{2}\right)  \right\}, \\
     \hat{d}_{t_1k_1t_2k_2}^{(x,z)} &= \frac{1}{n_z} \sum_{j=1}^{n_z} \left\{ \left(\frac{1}{n_x} \sum_{i=1}^{n_x} I(x_{it_1k_1} < z_{jt_1k_1}) - \frac{1 + \hat{\theta}_{t_1k_1}^{(x,z)}}{2}\right) 
    \left(\frac{1}{n_x} \sum_{i=1}^{n_x} I(x_{it_2k_2} < z_{jt_2k_2}) - \frac{1 + \hat{\theta}_{t_2k_2}^{(x,z)}}{2}\right)  \right\}, \\
    \hat{c}_{t_1k_1t_2k_2}^{(x,y,z)} &= \frac{1}{n_x} \sum_{i=1}^{n_x} \left\{ \left(\frac{1}{n_y} \sum_{j=1}^{n_y} I(y_{jt_1k_1} < x_{it_1k_1}) - \frac{1 - \hat{\theta}_{t_1k_1}^{(x,y)}}{2}\right)
    \left(\frac{1}{n_z} \sum_{l=1}^{n_z} I(z_{lt_2k_2} < x_{it_2k_2}) - \frac{1 - \hat{\theta}_{t_2k_2}^{(x,z)}}{2}\right)  \right\}.
\end{align*}
}
\end{theorem}

The proofs are given in the Section \ref{sec:proofs} (see \ref{proof1}, \ref{proof2} and \ref{proof3}).

\subsection{Extension to multi-arm LRST test statistics (more than 3 treatment arms) \label{subsec:22}}

Our methodology extends to multi-arm trials with more than three groups. In these cases, the test statistic involves the maximum of multiple dependent normal variables, for which a closed-form expression is unavailable. Therefore, the p-value cannot be derived exactly and must be obtained using numerical procedures.

Consider a scenario with four groups: one control group denoted by $\bm{X} = (\bm{X}_1^\top, \ldots, \bm{X}_T^\top)$, and three treatment groups denoted by $\bm{Y}_1$, $\bm{Y}_2$, and $\bm{Y}_3$, following the notation from Section~\ref{subsec:21}. In this setting, we define three relative treatment effect functions: $\theta_{tk}^{(x,y_1)}$, $\theta_{tk}^{(x,y_2)}$, and $\theta_{tk}^{(x,y_3)}$. The null hypothesis is

\begin{equation}
    H_0: \bar\theta^{(x,y_1)} = \bar\theta^{(x,y_2)} = \bar\theta^{(x,y_3)} = 0 \quad \textrm{ vs } \quad H_1: \bar{\theta}^{(x,y_1)} > 0 \text{ or } \bar{\theta}^{(x,y_2)}>0 \text{ or } \bar{\theta}^{(x,y_3)}>0,
% \, \text{not}\, H_0,
\end{equation}
where the average is taken over all $t$ and $k$. Rejecting the null hypothesis indicates that at least one treatment group has a significant effect, while non-rejection suggests no significant evidence of treatment effect in any group. 

Let $N = n_x + \sum_{i=1}^3 n_{y_i}$, where $n_x$ is the sample size for the control group, and $n_{y_i}$ is the sample size for the $i$th treatment group ($i=1, 2, 3$). Define $N_i=n_x+n_{y_i}$ for each treatment group $i$ and the corresponding ratios as $\lambda_{x} = \frac{n_x}{N}$, $\lambda_{y_i} = \frac{n_{y_i}}{N}$, $\lambda_{x,y_i} = \frac{n_x}{n_{y_i}}$, and $\lambda_x^{y_i} = \frac{n_x}{N_i}$. Let $\bm{R}_{\textnormal{diff}}^{x, y_1, y_2, y_3} = \left(\bm{R}_{\textnormal{diff}}^{(x,y_1)}, \bm{R}_{\textnormal{diff}}^{(x,y_2)}, \bm{R}_{\textnormal{diff}}^{(x,y_3)}\right)$, with each component defined as before. The covariance matrix is:

\[
\Sigma = 
\begin{pmatrix}
\Sigma^{(x,y_1)} & \Sigma^{(x,y_1,y_2)} & \Sigma^{(x,y_1,y_3)} 
% & \Sigma_{1,4}(\bm{X}, \bm{Y}_1, \bm{X}_4) 
\\
& \Sigma^{(x,y_2)} & \Sigma^{(x,y_2,y_3)}  
% & \Sigma_{2, K-1}(\bm{X}_1, \bm{X}_3, \bm{X}_k) 
\\
& & \Sigma^{(x,y_3)} 
% & \Sigma_{3, 4}(\bm{X}, \bm{Y}_3, \bm{X}_{K}) 
\\
% & & \vdots & & \\
% & & \cdots & & \Sigma_{K-1, K-1}(\bm{X}_1, \bm{X}_K)
\end{pmatrix},
\]

where, for $i=1,2,3$, 

\begin{align*}
    \Sigma^{(x,y_i)} &= \left( \lambda_{x} + \lambda_{y_i} \right)\frac{1}{K^2} \sum_{k_1, k_2 = 1}^K \left( \left(1 + \frac{1}{\lambda_{x}^{y_i}}\right)c_{t_1k_1t_2k_2}^{(x,y_i)} + \left(1 + \lambda_{x}^{y_i} \right)d_{t_1k_1t_2 k_2}^{(x,y_i)} \right), \\
    \Sigma^{(x,y_i,y_j)} &= \frac{1}{K^2} \sum_{k_1,k_2=1}^K \left( 1 + \frac{n_{y_i}}{n_x}\right)(\lambda_x + \lambda_{y_j}) c_{t_1k_1t_2k_2}^{(x,y_i,y_j)}.
\end{align*}

The test statistic is similar to \eqref{maxTest}:
\begin{equation}
    \txr{\mathcal{T}_{LRST}^{(x,y_1,y_2,y_3)}} = T^{-1}max \left\{\frac{\bar{R}_{y_1\cdot\cdot\cdot}^{(x,y_1)} - \bar{R}_{x\cdot\cdot\cdot}^{(x,y_1)}}{\sqrt{N J^T \hat{\Sigma}^{(x,y_1)}J}}, \frac{\bar{R}_{y_2\cdot\cdot\cdot}^{(x,y_2)} - \bar{R}_{x\cdot\cdot\cdot}^{(x,y_2)}}{\sqrt{NJ^T\hat{\Sigma}^{(x,y_2)}J}}, \frac{\bar{R}_{y_3\cdot\cdot\cdot}^{(x,y_3)} - \bar{R}_{x\cdot\cdot\cdot}^{(x,y_3)}}{\sqrt{NJ^T\hat{\Sigma}^{(x,y_3)}J}}\right\}.
\end{equation}

The distribution of the $\max$ test statistic is of the form:

\begin{align*}
     P(\txr{\mathcal{T}_{LRST}^{(x,y_1,y_2,y_3)}} \leq v) = \int_{u_1 = -\infty}^v \int_{u_2 = -\infty}^v \int_{u_3 =-\infty}^v (2 \pi)^3 det(\Sigma)^{-1/2} exp\left(-\frac{1}{2} \bm{u}^T \Sigma^{-1} \bm{u}\right)d\bm{u}, 
\end{align*}
where $\bm{u} = \{u_1, u_2, u_3\}$. The expression can be used to numerically obtain the p-value, or we can use Monte Carlo simulation to obtain the p-value numerically. 

\begin{remark}
The procedure can be generalized to multi-arm trials with any number of arms. When the number of arms exceeds three, numerical procedures are required to obtain the p-value since a closed-form expression of the test statistic's distribution is unavailable. For a trial with $A+1$ arms, including one control and $A$ treatment arms, the density of the test statistic $\mathcal{T}_{LRST}^{(x, \bm{y})}$, where $\bm{y} = (y_1, y_2, \ldots, y_A)$, corresponds to the density of the maximum of $A$ dependent normal random variables and is given by:

\begin{align*}
    P(\txr{\mathcal{T}_{LRST}^{(x, \bm{y})}} \leq v) = \int_{u_1 = -\infty}^v \ldots \int_{u_A = -\infty}^v (2 \pi)^A det(\Sigma)^{-1/2} exp\left(-\frac{1}{2} \bm{u}^T \Sigma^{-1} \bm{u}\right)d\bm{u}, 
\end{align*}
where the covariance matrix $\Sigma$ has the following elements:

\[
\Sigma = 
\begin{pmatrix}
\Sigma^{(x,y_1)} & \Sigma^{(x,y_1,y_2)} & \Sigma^{(x,y_1,y_3)} & \cdots & \Sigma^{(x,y_1,y_A)} \\
& \Sigma^{(x,y_2)} & \Sigma^{(x,y_2,y_3)} & \cdots & \Sigma^{(x,y_2,y_A)} \\
& & \Sigma^{(x,y_3)} & \cdots & \Sigma^{(x,y_3,y_A)} \\
& & \vdots & & \\
& & \cdots & & \Sigma^{(x,y_A)}
\end{pmatrix}.
\]

The covariance matrices can be estimated using the procedure outlined earlier. Specifically, the elements of the covariance matrix $\Sigma$ for $t_1, t_2 = 1, \ldots, T$ can be obtained as follows:

\begin{align*}
  \Sigma^{(x,y_i)}_{t_1,t_2} &= \left( \lambda_{x} + \lambda_{y_i} \right)\frac{1}{K^2} \sum_{k_1, k_2 = 1}^K \left( \left(1 + \frac{1}{\lambda_{x}^{y_i}}\right)c_{t_1 k_1 t_2 k_2}^{(x,y_i)} + \left(1 + \lambda_{x}^{y_i} \right)d_{t_1 k_1 t_2 k_2}^{(x,y_i)} \right),  \quad i=1,\ldots,A, \\
  \Sigma^{(x,y_i,y_j)} &= \frac{1}{K^2} \sum_{k_1,k_2=1}^K \left( 1 + \frac{n_{y_i}}{n_x}\right)(\lambda_x + \lambda_{y_j}) c_{t_1k_1t_2k_2}^{(x,y_i,y_j)},  \quad i,j=1,\ldots,A.
\end{align*}
The p-value for the test can be computed numerically, as in the four-arm case.
\end{remark}

\section{Simulation \label{sec:sim}}

\subsection{Simulation setting}

We conducted a comprehensive simulation study to evaluate the multi-arm LRST test's ability to control Type I error rates and detect significant treatment effects. The study design closely follows the Bapi 301 clinical trial (as described in Section~\ref{sec:realData}), utilizing the same primary endpoints ($K=2$ endpoints): the 11-item cognitive subscale of the Alzheimer's Disease Assessment Scale (ADAS-cog11) and the Disability Assessment for Dementia (DAD). In these simulations, higher ADAS-cog11 scores indicate greater cognitive impairment, while higher DAD scores indicate better cognitive function. Due to the opposing directions of these endpoints, negative values were used for the mean change of ADAS-cog11. 

We maintained similar sample sizes, trial progression patterns, duration, and assessment schedules ($T=6$ time points) as in the Bapi 301 study. To simulate the placebo arm data, we derived the mean changes from baseline to subsequent visits for ADAS-cog11 and DAD scores, along with their standard deviations (SD), based on values from the BAPI 301 study (see Table~\ref{tab:sdBAPI}). To account for within-subject temporal variability, we modeled random effects using an AR(1) structure with a lag-1 autocorrelation coefficient of 0.6. Additionally, we introduced inter-outcome correlation by setting the between-outcome correlation coefficient to 0.5. Each simulation scenario involved 1,000 simulations to calculate the empirical Type I error rates and power, \txr{and a nominal significance level of $0.05$ was used}.

\begin{table}[htbp] \footnotesize
    \centering
    \begin{tabular}{ccccccc} \hline
        Visit Week & 13 & 26 & 39 & 52 & 65 & 78  \\ \hline
         ADAS-cog11 & $0.601_{5.437}$ & $2.041_{5.813}$ & $3.139_{7.201}$ & $4.297_{8.151}$ & $5.643_{8.507}$ & $6.567_{9.511}$  \\ 
         DAD & $-1.740_{12.05}$ & $-3.539_{12.918}$ & $-6.719_{13.797}$ & $-9.420_{16.649}$ & $-11.287_{17.253}$ & $-12.958_{19.806}$ \\ \hline
    \end{tabular}
    \caption{Mean changes from baseline to each visit for ADAS-cog11 and DAD scores, with standard deviations in subscripts. Abbreviations: ADAS-cog11, Alzheimer's Disease Assessment Scale-cognitive subscale (11 items); DAD, Disability Assessment for Dementia.}
    \label{tab:sdBAPI}
\end{table}

\subsection{Type I error} 
To evaluate Type I error control, we conducted simulations under the null hypothesis, assuming no treatment effects across both outcomes and all visits for all treatment groups. Table~\ref{tab:typeI} presents the empirical Type I error rates of the multi-arm LRST at a nominal level of $\alpha=0.05$ across different sample sizes and varying numbers of treatment arms. The results show that the Type I error rates consistently remain close to the nominal level of 0.05, regardless of the number of arms or the sample size. This consistency highlights the robust error control and reliability of the proposed method in multi-arm trials.

\begin{table}[htbp]
    \centering
    \begin{tabular}{c rrrrr} \hline
       $n_x $  & 3-arm (N) & 4-arm (N) & 5-arm (N) & 6-arm (N) & 7-arm (N) \\ \hline
        200  & 0.047 (467) & 0.045 (600) & 0.042 (734) & 0.048 (867) & 0.056 (1000) \\
        300 & 0.045 (700) & 0.056 (900) & 0.055 (1100) & 0.049 (1300) & 0.047 (1500) \\
        400 & 0.048 (934)  & 0.052 (1200) & 0.049 (1467) & 0.051 (1734) & 0.056 (2000) \\
        500 & 0.042 (1167) & 0.048 (1500) & 0.050 (1833) & 0.051 (2167) & 0.047 (2500) \\ \hline
    \end{tabular}
    \caption{Type I error rates of the multi-arm LRST at $\alpha = 0.05$ for various sample sizes and numbers of treatment arms. $n_x$, the sample size in the control group; $N$, the total sample size; A, the number of arms. The allocation proportion is maintained at $3:2:2:\ldots:2$, indicating that the control arm's sample size is $n_x=\frac{3}{3 + 2(A-1)}N$, while each treatment arm has a sample size of $\frac{2}{3 + 2(A-1)}N$. The results are based on 1,000 simulations.}
    \label{tab:typeI}
\end{table}

\subsection{Power assessment in three-arm trials \label{sec:sim1}}
\begin{figure}[htbp]
    \centering
    \includegraphics[width = \textwidth]{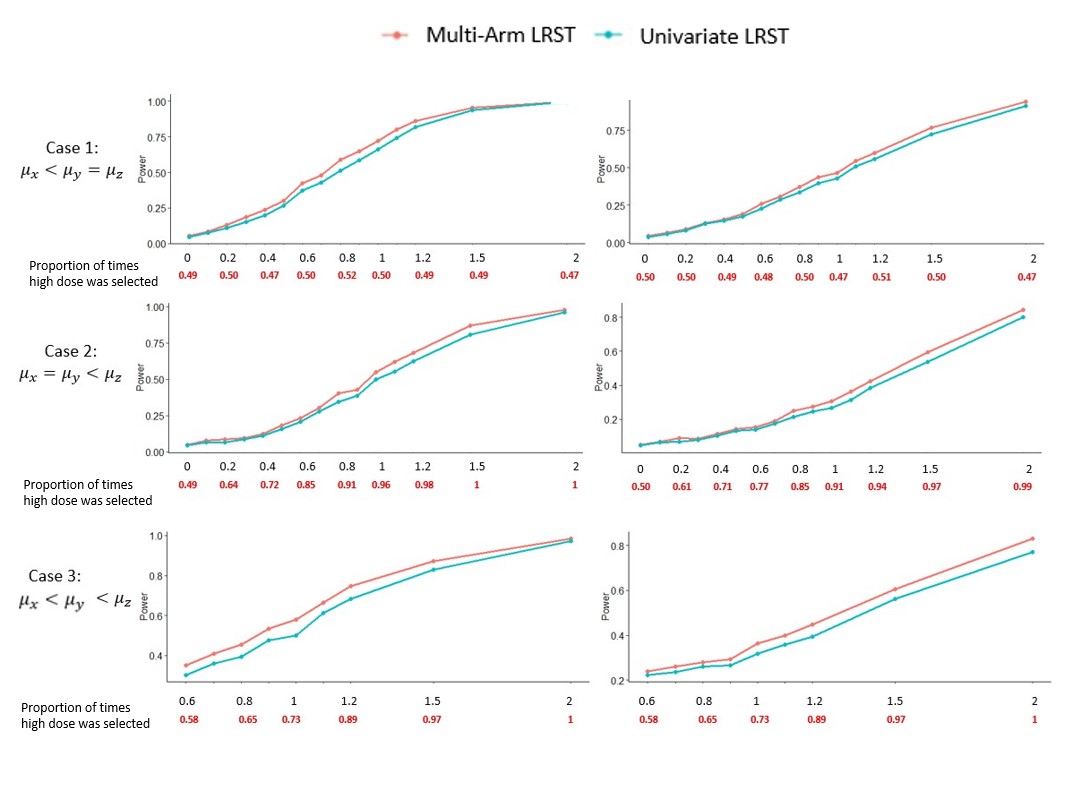}
    \caption{Power curves for Cases 1, 2, and 3 (rows 1, 2, and 3) under two sample sizes: $(373, 238, 226)$ (left) and $(180, 120, 120)$ (right) with a $3:2:2$ allocation ratio. Treatment mean curves are denoted as $\mu_x$ (control), $\mu_y$ (low-dose), and $\mu_z$ (high-dose). The x-axis shows the effect size multiplier $\alpha$, with the proportion of times the high-dose group was selected as having the maximum effect displayed below. Case 1: $\alpha$ varied from $0$ to $2$ for both $\mu_y$ and $\mu_z$. Case 2: $\alpha$ was $0$ for $\mu_y$ and varied for $\mu_z$. Case 3: $\alpha$ was $0.5$ for $\mu_y$ and ranged from $0.6$ to $2$ for $\mu_z$.}  \label{fig:powerTotal}
\end{figure}

We conducted a simulation study to evaluate the efficacy of the multi-arm LRST in detecting changes in the mean function across three arms: control, low-dose ($0.5$ mg/kg), and high-dose ($1.0$ mg/kg). The study utilized data from the BAPI 301 study, with mean curves denoted as $\mu_x$ (control), $\mu_y$ (low-dose), and $\mu_z$ (high-dose). Two total sample sizes were considered, maintaining the original allocation ratio of 3:2:2: 373 for control, 238 for low-dose, and 226 for high-dose (Figure~\ref{fig:powerTotal}, left panels), and a smaller sample size group of $(180, 120, 120)$ (Figure~\ref{fig:powerTotal}, right panels). 

Power was assessed for the effect sizes specified in the BAPI 301 study protocol \cite{salloway2014two}, which are $2.65$ for the ADAS-cog11 score and $6.56$ for the DAD score. We calculated power across a range of multipliers for these effect sizes, represented as $\alpha \times (2.65, 6.56)$. For example, with $\alpha = 0.5$, the effect sizes are $1.325$ for ADAS-cog11 and $3.28$ for DAD. For comparison, we also applied the univariate LRST \cite{xu2025novel} with a Bonferroni correction (referred to as the univariate LRST), rejecting the null hypothesis if at least one of the p-values from the two univariate LRST tests was less than $0.025$ ($0.05/2$).

Three distinct cases were evaluated, each with two different  sample sizes:
\begin{itemize}
\item Case 1: $\mu_x < \mu_Y = \mu_Z$, indicating equal positive effects in both dose groups. The effect size multiplier $\alpha$ varied from $0$ to $1.5$ in increments of $0.1$, with an additional value at $2$ for both the low-dose and high-dose groups. 

\item Case 2: $\mu_x = \mu_Y < \mu_Z$, suggesting no effect in the low-dose group and a positive effect in the high-dose group. The effect size multiplier was set to 0 for the low-dose group, while the high-dose group's multiplier varied as in Case 1. 

\item Case 3: $\mu_x < \mu_Y < \mu_Z$, indicating unequal positive effects in both dose groups. The multiplier was set to 0.5 for the low-dose group, and ranged from $0.6$ to $1.2$ in increments of 0.1, with additional values at $1.5$ and $2$ for the high-dose group.  
\end{itemize}

The proportion of times the high-dose group was selected as having the maximum treatment effect is displayed below the x-axis for each effect size multiplier $\alpha$ in Figure~\ref{fig:powerTotal}. In Case 1, where both dose groups have equal positive effects, the power curves (Figure~\ref{fig:powerTotal}, first row) demonstrate that the multi-arm LRST consistently outperforming the univariate LRST in power. As expected, given the equal effect sizes, both doses are equally likely to be selected as the most efficacious dose when the null hypothesis is rejected, with proportions close to $0.5$. 

In Case 2, where only high-dose group has a positive effect, the power curves (Figure~\ref{fig:powerTotal}, second row) shows a similar trend to Case 1, with the multi-arm LRST outperforming the univariate LRST as the effect size multiplier $\alpha$ increases. The proportion of times the high-dose group was selected increases, approaching $1$. In Case 3, where positive effects are present in both treatment doses with increasing magnitude, the power curves (Figure~\ref{fig:powerTotal}, third row) suggest that the multi-arm LRST consistently outperforms the univariate LRST. The power curves increases with the effect size and eventually converging to $1$. Additionally, the proportion of times the high-dose group was selected increases with the effect size, approaching $1$ under large effect sizes.

\subsection{Power assessment in four-arm trials \label{sec:sim2}}
\begin{figure}[htbp]
{
    \centering
    \includegraphics[width = \textwidth]{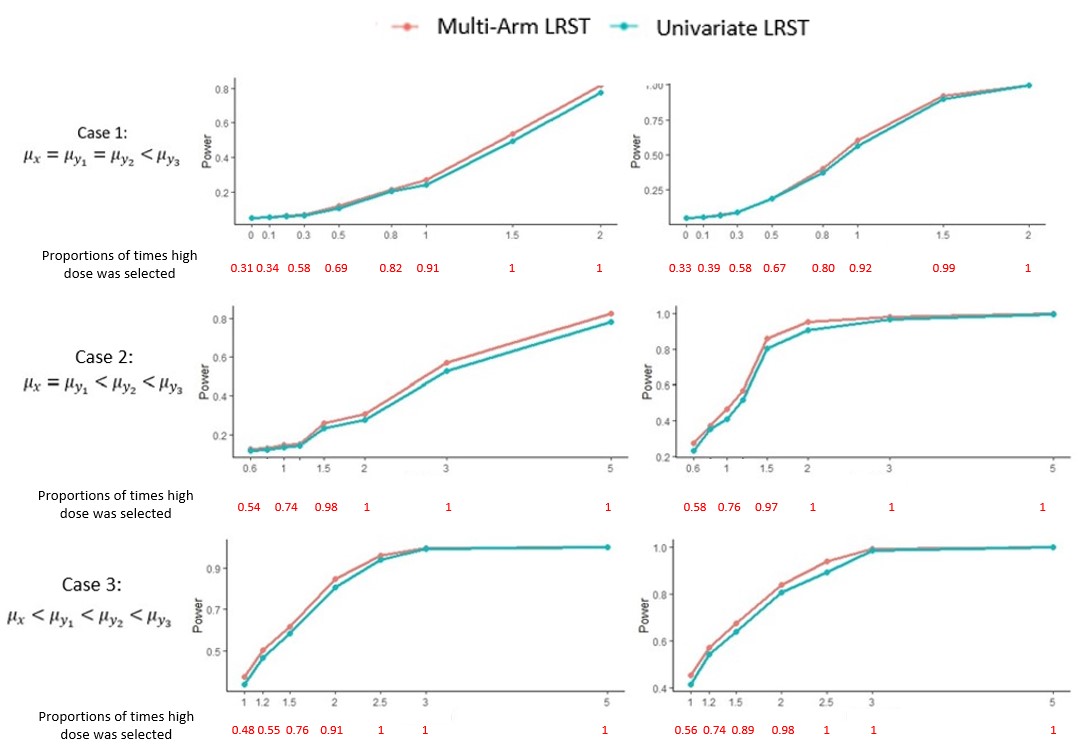}
    \caption{Power curves for Cases 1, 2, and 3 (rows 1, 2, and 3) in the four-arm trial, with sample sizes $(200, 134, 134, 134)$ (left) and $(300, 200, 200, 200)$ (right) under a $3:2:2:2$ allocation ratio. Treatment mean curves are denoted as $\mu_x$ (control), $\mu_{y_1}$ (low-dose), $\mu_{y_2}$ (mid-dose), and $\mu_{y_3}$ (high-dose). The x-axis shows the effect size multiplier $\alpha$, with the proportion of times the high-dose group was selected as having the maximum effect displayed below. Case 1: $\alpha$ was $0$ for $\mu_{y_1}$ and $\mu_{y_2}$, and varied for $\mu_{y_3}$. Case 2: $\alpha$ was $0$ for $\mu_{y_1}$, $0.5$ for $\mu_{y_2}$, and varied for $\mu_{y_3}$. Case 3: $\alpha$ was $0.5$ for $\mu_{y_1}$, $0.8$ for $\mu_{y_2}$, and varied for $\mu_{y_3}$.} 
    \label{fig:powerTotalfour}
    }
\end{figure}

We extended the simulation study to a four-arm trial by adding an additional treatment group, with treatment mean curves denoted as $\mu_{y_1}$ (low-dose), $\mu_{y_2}$ (mid-dose), and $\mu_{y_3}$ (high-dose). The effect multipliers for these curves were applied as in Section~\ref{sec:sim1}. Two sample sizes were considered: $(n_x = 200, n_{y_i} = 134)$ (Figure~\ref{fig:powerTotalfour}, left panels) and $(n_x = 300, n_{y_i} = 200)$ (Figure~\ref{fig:powerTotalfour}, right panels) for $i=1,2,3$, with the allocation ratio of 3:2:2:2.  

We compared the multi-arm LRST against the univariate LRST, similar to the three-arm trial. The proportion of times the high-dose group was selected as having the maximum treatment effect is displayed below the x-axis for each effect size multiplier $\alpha$ in Figure~\ref{fig:powerTotalfour}. 

Three distinct cases were evaluated, each with two different sample sizes:
\begin{itemize}
    \item Case 1: $\mu_{x} = \mu_{y_1} = \mu_{y_2} < \mu_{y_3} $ indicating a positive effect only in the high-dose group. The effect size multiplier $\alpha$ was set to 0 for the low-dose and mid-dose groups, while the high-dose group's multiplier varied across values of 0, 0.1, 0.2, 0.3, 0.5, 0.8, 1, 1.5, and 2. 

    \item Case 2: $\mu_{x} = \mu_{y_1} < \mu_{y_2} < \mu_{y_3}$ indicating unequal positive effects in the mid-dose and high-dose groups. The multiplier $\alpha$ was set to 0 for the low-dose group and 0.5 for the mid-dose group, and varied across  0.6, 0.8, 1, 1.2, 1.5, 2, 3, and 5 for the high-dose group.

    \item Case 3: $\mu_{x} < \mu_{y_1} < \mu_{y_2} < \mu_{y_3} $ indicating unequal positive effects across all treatment doses. The multiplier $\alpha$ was set to 0.5 for the low-dose group, 0.8 for the mid-dose group, and varied across 1, 1.2, 1.5, 2, 2.5, 3, and 5 for the high-dose group. 
\end{itemize}

In Case 1, where only the high-dose group shows a positive effect, the power curves (Figure~\ref{fig:powerTotalfour}, first row) demonstrate that the multi-arm LRST outperforms the univariate LRST, with power increasing as the effect size multiplier $\alpha$ increases. As $\alpha$ increases, the proportion of times the high-dose group was selected as the most efficacious dose increases, approaching 1. In Case 2, involving positive effects in both the mid-dose and high-dose groups but to different extents, the power curves (Figure~\ref{fig:powerTotalfour}, second row) show that the multi-arm LRST consistently outperforms the univariate LRST. As $\alpha$ increases, the high-dose group was more frequently selected as the target dose, with proportions increasing toward $1$ as the effect size grows. In Case 3, where positive effects are present across all treatment doses with increasing magnitude, the power curves (Figure~\ref{fig:powerTotalfour}, third row) illustrate the superior performance of the multi-arm LRST compared to the univariate LRST. The power increases with the effect size, and the proportion of times the high-dose group is selected also increases, approaching $1$ for large effect sizes.

In conclusion, the multi-arm LRST consistently outperformed the univariate LRST in power across all cases, particularly in detecting treatment effects across multiple doses. The multi-arm LRST effectively identified the most effective treatment dose with the maximum effect, even when effects were unequal across different dose groups, demonstrating robust performance across various sample sizes and configurations, highlighting its reliability as a powerful tool for multi-arm clinical trials.

\subsection{Runtime performance \label{sec:runTime}}

The runtime performance of the multi-arm LRST, summarized in Table~\ref{tab:runTime} highlights its computational efficiency across varying sample sizes and numbers of treatment arms. Despite the increase in sample size ($N$) and complexity as the number of arms grows, the method demonstrates scalable and rapid computation. For example, with a control group size of $n_x=200$, the runtime remains efficient at just 0.505 seconds for a 3-arm trial ($N=467$) and increases modestly to 0.991 seconds for a 7-arm trial ($N=1000$). Similarly, when $n_x=500$, the runtime increases from 0.796 seconds for a 3-arm trial ($N=1167$) to 1.747 seconds for a 7-arm trial ($N=2500$), illustrating only moderate growth relative to the increased complexity and sample size. This scalability demonstrates the suitability of the multi-arm LRST for real-world clinical trials where rapid computation is critical. Furthermore, the method achieves this efficiency while handling multivariate longitudinal data and avoiding the computational burdens associated with stringent distributional assumptions or multiplicity adjustments common in traditional approaches. The results establish the multi-arm LRST as a computationally robust and practical tool for analyzing modern clinical trial data.

%The table \ref{tab:runTime} highlights the runtime performance of the multi-arm LRST, showcasing its computational efficiency across various sample sizes and numbers of treatment arms. Even as the total sample size (N) grows substantially with increasing arms, the method maintains remarkably fast runtimes. For instance, with a control group size of $200$ ($n_x = 200$, the runtime is only $0.505$ seconds for a 3-arm trial and $0.991$ seconds for a 7-arm trial. Similarly, for $n_x = 500$, the runtime increases modestly from $0.796$ seconds for a 3-arm trial to $1.747$ seconds for a 7-arm trial, despite the significant increase in complexity and sample size. This demonstrates that the multi-arm LRST is highly scalable and suitable for real-world clinical trial applications where rapid computation is essential. The computational speed is especially notable given the method's ability to handle multivariate longitudinal data while avoiding the need for stringent distributional assumptions or multiplicity adjustments, a common computational burden in traditional methods.

\begin{table}[htbp]
    \centering
    \begin{tabular}{c rrrrr} \hline
       $n_x $  & 3-arm ($N$) & 4-arm ($N$) & 5-arm ($N$) & 6-arm ($N$) & 7-arm ($N$) \\ \hline
        200  & 0.505 (467) & 0.614 (600) & 0.723 (734) & 0.870 (867) & 0.991 (1000) \\
        300 & 0.604 (700) & 0.758 (900) & 0.890 (1100) & 1.083 (1300) & 1.226 (1500) \\
        400 & 0.682 (934)  & 0.898 (1200) & 1.065(1467) & 1.293 (1734) & 1.507 (2000) \\
        500 & 0.796 (1167) & 1.078 (1500) & 1.300(1833) & 1.513 (2167) & 1.747 (2500) \\ \hline
    \end{tabular}
    \caption{Runtime (in seconds) for multi-arm LRST for various sample sizes and numbers of treatment arms. $n_x$: the control group sample size; $N$: the total sample size; A: the number of arms. The allocation proportion is maintained at $3:2:2:\ldots:2$, indicating that the control arm's sample size is $n_x=\frac{3}{3 + 2(A-1)}N$, while each treatment arm has a sample size of $\frac{2}{3 + 2(A-1)}N$.}
    \label{tab:runTime}
\end{table}

\section{Real Data Analysis \label{sec:realData}}

The Bapineuzumab (Bapi) 301 trial, detailed in \cite{salloway2014two}, was a double-blind, randomized, placebo-controlled, phase 3 clinical trial designed to evaluate the efficacy of intravenous Bapineuzumab in patients with mild-to-moderate Alzheimer’s disease (AD) who were noncarriers of the apolipoprotein E (APOE) $\epsilon$4 allele. Conducted across 191 sites across 29 countries from December 2007 through April 2012, participants were randomized in a 3:2:2 ratio into placebo, low-dose (0.5 mg/kg), and high-dose (1.0 mg/kg) Bapi groups. Assessments occurred at baseline and every 13 weeks for a total of 78 weeks ($t=13$, $26$, $39$, $52$, $65$, $78$ weeks with $T=6$).

\begin{figure}[htbp]
    \centering
    \includegraphics[width = \textwidth]{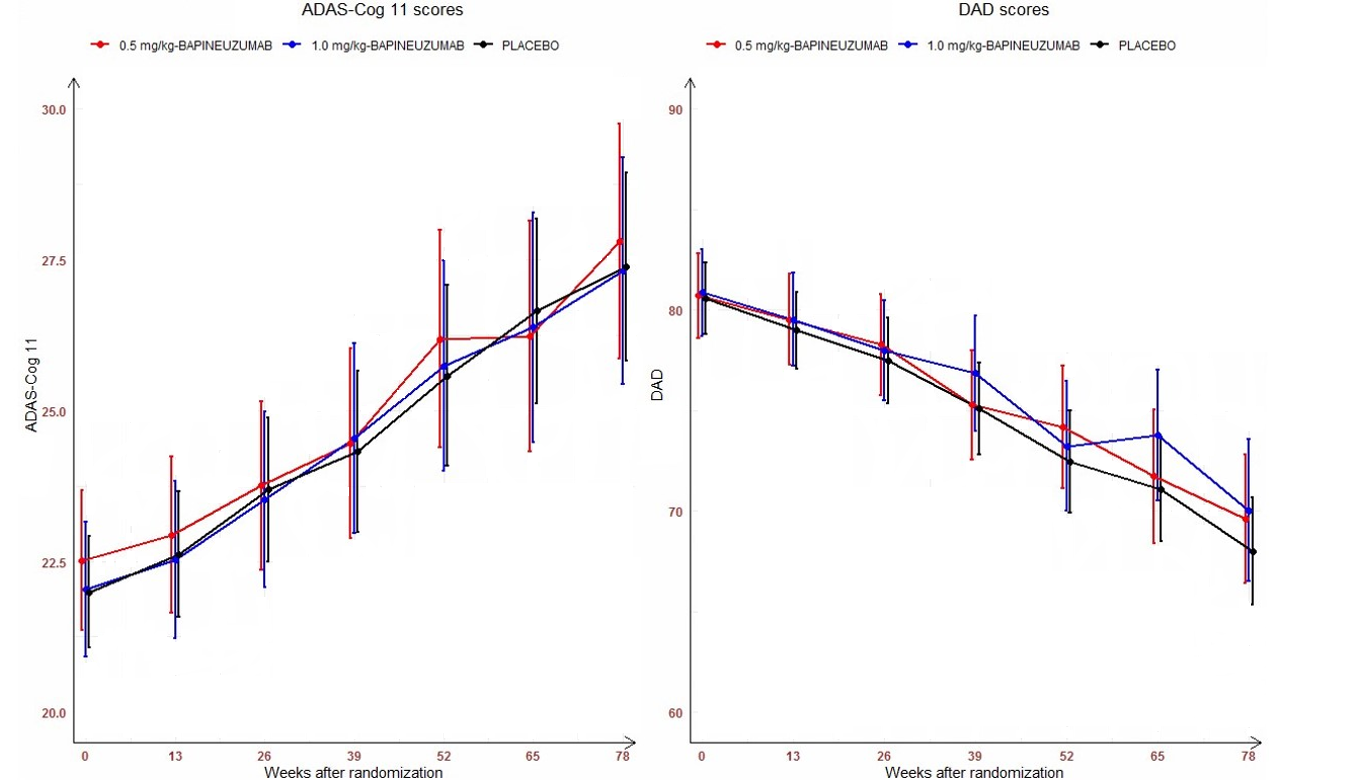}
    \caption{Mean changes from baseline to week 78 in scores on ADAS-cog11 (left panel, with scores ranging from 0 to 70 and higher scores indicating greater impairment) and DAD (right panel, with scores ranging from 0 to 100 and higher scores indicating less impairment) for the placebo (black), Bapi low-dose (red), and Bapi high-dose (blue) groups in the Bapineuzumab 301 Trial. Vertical lines are 95\% confidence intervals. Abbreviation: ADAS-cog11, Alzheimer’s Disease Assessment Scale - cognitive subscale (11 items); DAD, Disability Assessment for Dementia. \label{fig:adas-dad-1}}
\end{figure}

The co-primary endpoints were ADAS-cog11 and DAD, requiring significance in both to confirm Bapineuzumab's efficacy. ADAS-cog11, a comprehensive tool consisting of 11 items, evaluates cognitive functions across various domains including memory, language, praxis, and orientation, with scores ranging from 0 to 70, where higher scores greater cognitive impairment. DAD, based on caregiver assessments over 40 items, measured the patient’s autonomy in daily tasks, such as personal hygiene and meal preparation, with scores ranging from 0 to 100, where higher scores reflect greater independence. This analysis included only patients who had complete observations of ADAS-cog11 and DAD throughout the trial, resulting in sample sizes of 324 (placebo), 196 (low dose), and 205 (high dose), with the 3:2:2 allocation ratio and total sample size $N=725$. Baseline characteristics are presented in Table~\ref{characTab}, which shows that the groups were well-balanced in terms of age, gender distribution, and baseline ADAS-cog11 and DAD scores. Figure~\ref{fig:adas-dad-1} illustrates the mean trajectories of ADAS-cog11 and DAD scores over 78 weeks, with the increase in ADAS-cog11 scores reflecting cognitive decline and the decrease in DAD scores indicating reduced functioning, consistent across all groups. Notably, there is no obvious treatment efficacy observed in the Bapi dose groups compared to the placebo, as the trajectories are similar across all treatment arms.

\begin{table}[htbp]
\begin{center}
{
\centering
\begin{tabular}{lrrr}
\hline
\cline{2-4} 
\multirow{-2}{*}{{\color[HTML]{333333} Characteristic}} & \multicolumn{1}{c}{\begin{tabular}[c]{@{}c@{}}Control\\      (N = 324)\end{tabular}} & \multicolumn{1}{c}{\begin{tabular}[c]{@{}c@{}}Bapineuzumab\\ 0.5 mg/kg\\ (N = 196)\end{tabular}} & \begin{tabular}[c]{@{}c@{}}Bapineuzumab\\  1.0 mg/kg   \\ (N   = 205)\end{tabular} \\ \hline
Age, y, mean (SD) & 61.29 (2.96) & 62.65 (12.53) & 62.73 (11.89) \\
Female sex, n (\%) & 167 (51.50) & 103 (52.50) & 110 (53.70) \\
ADAS-cog11, mean (SD) & 23.79 (12.78) & 23.66 (12.33) & 23.66 (12.44) \\
DAD, mean (SD) & 76.50 (22.17) & 77.67 (20.76) & 78.83 (20.99) \\ \hline
\end{tabular}
\caption{Baseline characteristics for the BAPI 301 study. SD, standard deviation; ADAS-cog11, Alzheimer’s Disease Assessment Scale - cognitive subscale (11 items); DAD, Disability Assessment for Dementia.}
\label{characTab}
}
\end{center}
\end{table}

For the analysis, we combined the outcome vector from the placebo group $(x_{1tk}, \ldots, x_{n_xtk})$ with that from the low-dose group $(y_{1tk}, \ldots, y_{n_ytk})$, and separately with the high-dose group $(z_{1tk}, \ldots, z_{n_ztk})$, ranking them accordingly to compute the test statistic. The rank difference vectors were calculated as $\bar{R}_{y\cdot\cdot\cdot}^{(x,y)} - \bar{R}_{x\cdot\cdot\cdot}^{(x,y)} = 9.689$ and $\bar{R}_{z\cdot\cdot\cdot}^{(x,z)} - \bar{R}_{x\cdot\cdot\cdot}^{(x,z)} = 66.115$, with $\sqrt{N J^T \hat{\Sigma}^{(x,y)} J} = 11.228$ and $\sqrt{N J^T \hat{\Sigma}^{(x,z)} J} = 11.363$. Consequently, we obtained $\frac{\bar{R}_{y\cdot\cdot\cdot}^{(x,y)} - \bar{R}_{x\cdot\cdot\cdot}^{(x,y)}}{\sqrt{N J^T \hat{\Sigma}^{(x, y)}J}} = 0.86$ and $\frac{\bar{R}_{y\cdot\cdot\cdot}^{(x,z)} - \bar{R}_{x\cdot\cdot\cdot}^{(x,z)}}{\sqrt{N J^T \hat{\Sigma}^{(x, z)}J}} = 5.82$, leading to the final test statistic $\mathcal{T}_{LRST}^{x,y,z} = 5.82/6 = 0.970$. The p-value for the multi-arm LRST was $0.253$, indicating no significant evidence of treatment efficacy for either outcome. Consequently, further separate testing of treatment efficacy on the two outcomes was unnecessary. However, we also performed the univariate LRST for placebo vs. low-dose and placebo vs. high-dose, yielding p-values of 0.455 and 0.201, respectively, both of which were well above the Bonferroni-corrected threshold of 0.025.

\section{Conclusion}

This manuscript presents the multi-arm Longitudinal Rank-Sum Test (LRST), a novel U-statistics-based method designed for multi-arm clinical trials with multiple longitudinal primary outcomes. This non-parametric, rank-based method provides a robust alternative to traditional parametric approaches, which often rely on restrictive distributional assumptions. By extending the LRST \cite{xu2025novel} to accommodate multi-arm designs, our method offers a comprehensive solution for assessing treatment efficacy across multiple treatment arms and outcomes. Importantly, when the number of arms is two, the multi-arm LRST reduces to the regular LRST \cite{xu2025novel}, thus maintaining its simplicity while extending its applicability.

Our extensive simulation studies demonstrate that the multi-arm LRST not only maintains strong control over Type I error but also offers enhanced statistical power. Compared to the Bonferroni-corrected univariate LRST, our method consistently proves more efficient, making better use of the available clinical trial data.

The practical utility of the multi-arm LRST was further validated through its application to the Bapi 301 trial data. The analysis confirmed that Bapineuzumab did not have a statistically significant effect on the co-primary outcomes, ADAS-cog11 and DAD, reinforcing the drug's lack of efficacy in the studied population. This case study highlights the robustness of the multi-arm LRST in managing complex, real-world clinical trial data, establishing it as a valuable tool for future research, particularly in the field of neurodegenerative diseases and beyond.

In this article, we focused on directional hypotheses ($\bar{\theta}^{(x,y)} > 0$ or $\bar{\theta}^{(x,z)}>0$), reflecting the primary goal of many clinical trials to establish the superiority of treatment over control. The LRST framework can also accommodate non-directional hypotheses ($\bar{\theta}^{(x,y)} \neq 0$ or $\bar{\theta}^{(x,z)} \neq 0$) using an $\ell_2$-based test statistic as an alternative to the current $\ell_\infty$-based test statistic. For example, instead of $\mathcal{T}_{LRST}^{(x,y,z)}$ as defined in \eqref{maxTest}, the following $\ell_2$-based test statistic can be proposed:
\begin{equation*}
    \mathcal{T}_{\ell_2}^{(x,y,z)} = T^{-1}K^{-1} \left\{\left(\frac{\bar{R}^{(x,y)}_{y\cdot\cdot\cdot}-\bar{R}^{(x,y)}_{x\cdot\cdot\cdot}}{\sqrt{N J^T \hat{\Sigma}^{(x,y)} J}} \right)^2 +  \left(\frac{\bar{R}^{(x,z)}_{z\cdot\cdot\cdot}-\bar{R}^{(x,z)}_{x\cdot\cdot\cdot}}{\sqrt{NJ^T\hat{\Sigma}^{(x,z)} J}}\right)^2\right\},
\end{equation*}
which asymptotically converges to the sum of squares of dependent normal random variables. This sum can be expressed as a linear combination of independent non-central $\chi^2$ distributions. While we emphasized one-sided hypotheses relevant for efficacy testing in clinical trials, the $\ell_2$-based test offers a promising direction for non-directional hypotheses and merits further exploration.

\bibliographystyle{myjmva}
\bibliography{2ref}

\appendix

\section{Proofs}
\label{sec:proofs}
\subsection{Proof of Theorem \ref{thm1}\label{proof1}
}
\begin{proof}
    We begin by noting from \cite{xu2025novel} that $\frac{1}{\sqrt{N_1}} \bm{R}_{\textnormal{diff}}^{(x,y)} $ asymptotically follows $\mathcal{N}\left(\frac{\sqrt{N_1}}{2} \pmb{\theta}^{(x,y)}, \tilde{\Sigma}^{(x,y)}\right)$, where:
$$\tilde{\Sigma}^{(x,y)} = \frac{1}{K^2}\sum_{k_1,k_2 = 1}^K \left(\left(1 + \frac{1}{\lambda_x^y}\right)c_{t_1k_1t_2k_2}^{(x,y)} +  (1 + \lambda_x^y)d_{t_1k_1t_2k_2}^{(x,y)} \right).$$
Similarly, $\frac{1}{\sqrt{N_1}} \bm{R}_{\textnormal{diff}}^{(x,y)}$ asymptotically follows $\mathcal{N}\left(\frac{\sqrt{N_1}}{2} \pmb{\theta}^{(x,z)}, \tilde{\Sigma}^{(x,z)}\right)$, where:$$
\tilde{\Sigma}^{(x,z)} = \frac{1}{K^2}\sum_{k_1,k_2 = 1}^K \left(\left(1 + \frac{1}{\lambda_x^z}\right)c_{t_1k_1t_2k_2}^{(x,z)} +  (1 + \lambda_x^z)d_{t_1k_1t_2k_2}^{(x,z)} \right).
$$ 

Hence, by scaling, we find: $\frac{1}{\sqrt{N}}\bm{R}_{\textnormal{diff}}^{(x,y)}$ asymptotically follows $\mathcal{N}\left( \frac{(\lambda_x + \lambda_y)\sqrt{N}}{2} \pmb{\theta}^{(x,y)}, (\lambda_x + \lambda_y)\tilde{\Sigma}^{(x,y)} \right)$, and Similarly, $\frac{1}{\sqrt{N}}\bm{R}_{\textnormal{diff}}^{(x,z)}$ asymptotically follows $\mathcal{N}\left( \frac{(\lambda_x + \lambda_z)\sqrt{N}}{2} \pmb{\theta}^{(x,z)}, (\lambda_x + \lambda_z)\tilde{\Sigma}^{(x,z)} \right)$, since $\frac{N_1}{N} = \lambda_x + \lambda_y$. Hence, we can express:
$$\Sigma^{(x,y)} = (\lambda_x + \lambda_y) \tilde{\Sigma}^{(x,y)} \text{ and } \Sigma^{(x,z)} = (\lambda_x + \lambda_z) \tilde{\Sigma}^{(x,z)}. 
$$
Next, we compute the covariance between $R_{\textnormal{diff}, t_1}^{(x,y)}$ and $R_{\textnormal{diff}, t_2}^{(x,z)}$:
\begin{align*}
    Cov \left( R_{\textnormal{diff}, t_1}^{(x,y)}, R_{\textnormal{diff}, t_2}^{(x,z)} \right) &= Cov\left(\overline{R}_{y \cdot t_1 \cdot}^{(x,y)} - \overline{R}_{x \cdot t_1 \cdot}^{(x,y)}, \overline{R}_{z \cdot t_2 \cdot}^{(x,z)} - \overline{R}_{x \cdot t_2 \cdot}^{(x,z)} \right) \\
    &= \frac{N_1 N_2}{4K^2 n_x^2n_y n_z} \sum_{k_1,k_2=1}^K \sum_{i_1,i_2=1}^{n_x}\sum_{j_1=1}^{n_y}\sum_{j_2=1}^{n_z} Cov\Bigg\{\Big\{\mathbb{I}(x_{i_1 t_1 k_1} < y_{j_1 t_1 k_1}) \\
    &- \mathbb{I}(x_{i_1 t_1 k_1} > y_{j_1 t_1 k_1})\Big\},
    \Big\{\mathbb{I}(x_{i_2 t_2 k_2} < z_{j_2 t_2 k_2}) - \mathbb{I}(x_{i_2 t_2 k_2} > z_{j_2 t_2 k_2})\Big\}\Bigg\}  \\
    &= \frac{(n_x + n_y)(n_x + n_z)}{4K^2n_x^2n_yn_z} \sum_{k_1,k_2=1}^K \sum_{i_1,i_2=1}^{n_x}\sum_{j_1=1}^{n_y}\sum_{j_2=1}^{n_z} \zeta^{i_1,i_2,j_1,j_2}_{t_1,t_2,k_1,k_2}.
\end{align*}

Case 1: $i_1 \neq i_2$ and $j_1 \neq j_2$, then $\zeta^{i_1,i_2,j_1,j_2}_{t_1,t_2,k_1,k_2} = 0$

Case 2: $i_1 = i_2 = i$ and $j_1 \neq j_2$: 
\begin{align*}
    \zeta^{i,i,j_1,j_2}_{t_1,t_2,k_1,k_2} &=  Cov\Bigg\{\Big\{\mathbb{I}(x_{i_1 t_1 k_1} < y_{j_1 t_1 k_1}) - \mathbb{I}(x_{i_1 t_1 k_1} > y_{j_1 t_1 k_1})\Big\}, \Big\{\mathbb{I}(x_{i_2 t_2 k_2} < z_{j_2 t_2 k_2}) - \mathbb{I}(x_{i_2 t_2 k_2} > z_{j_2 t_2 k_2})\Big\}\Bigg\}  \\
    &= E\Bigg\{ \mathbb{I}(x_{it_1k_1} < y_{j_1t_1k_1} \quad \& \quad  x_{it_2k_2} < z_{j_2t_2k_2}) -  \mathbb{I}(x_{it_1k_1} < y_{j_1t_1k_1} \quad \& \quad x_{it_2k_2} > z_{j_2t_2k_2}) \\ 
    & - \mathbb{I}(x_{it_1k_1} > y_{j_1t_1k_1} \quad \& \quad x_{it_2k_2} < z_{j_2t_2k_2})  + \mathbb{I}(x_{it_1k_1} > y_{j_1t_1k_1} \quad \& \quad  x_{it_2k_2} > z_{j_2t_2k_2})\Bigg\}  - \theta_{t_1k_1}^{(x,y)}\theta_{t_2k_2}^{(x,z)} \\
    & = E \left\{ (1- 2 G_{t_1k_1}(X_{t_1k_1}))(1- 2 H_{t_2k_2}(X_{t_2k_2})) \right\} - \theta_{t_1k_1}^{(x,y)}\theta_{t_2k_2}^{(x,z)} \\
    & = Cov\left\{ (1- 2 G_{t_1k_1}(X_{t_1k_1})),(1- 2 H_{t_2k_2}(X_{t_2k_2})) \right\} \\&= 4 Cov \left(G_{t_1k_1}(X_{t_1k_1}), H_{t_2k_2}(X_{t_2k_2}) \right).
\end{align*}

Thus, assuming $n_{y \wedge z} = min(n_y, n_z)$:

\begin{align*}
   Cov \left\{R_{\textnormal{diff},t_1}^{(x,y)}, R_{\textnormal{diff},t_2}^{(x,z)}\right\} 
    & =  Cov \left\{  \overline{R}_{y\cdot t_1 \cdot}^{(x,y)} - \overline{R}_{x\cdot t_1 \cdot}^{(x,z)}, \overline{R}_{z\cdot t_2 \cdot}^{(x,z)} - \overline{R}_{x\cdot t_2 \cdot}^{(x,z)}\right\}  \\
    &= \frac{N_1 N_2}{4K^2 n_x^2n_y n_z} \sum_{k_1,k_2=1}^K \Bigg\{ 4 n_x \left(n_y n_z - n_{y \wedge z}\right) Cov \left(G_{t_1k_1}(X_{t_1k_1}), H_{t_2k_2}(X_{t_2k_2}) \right) 
    % + \\
    % &\quad\quad\quad 
    % 4 n_x(n_x-1)min(n_y,n_z) Cov\left\{F_{t_1k_1}(Y_{t_1k_1}, F_{t_2k_2}(Z_{t_2k_2})) \right\} 
    + n_x n_{y \wedge z}\zeta_{t_1k_1t_2k_2}^{iijj} \Bigg\}\\
    &= \frac{1}{K^2}\sum_{k_1, k_2 = 1}^K \left(\beta  Cov \left(G_{t_1k_1}(X_{t_1k_1}), H_{t_2k_2}(X_{t_2k_2}) \right)  
    % + \beta Cov\left\{F_{t_1k_1}(Y_{t_1k_1}, F_{t_2k_2}(Z_{t_2k_2})) \right\} 
    + \gamma/4 \zeta_{t_1k_1t_2k_2}^{iijj}\right),
    % &= \frac{1}{K^2}\sum_{k_1,k_2=1}^K \left\{ \left(1 + \frac{1}{\lambda}\right)c_{t_1k_1t_2k_2}^{X,Y,Z} + \left((1 + \lambda \right)d_{t_1k_1t_2k_2}^{X,Y,Z} \right\} + o(1)
\end{align*}

where:
\begin{align*}
    \beta &= \frac{N_1N_2}{n_x^2 n_y n_z}(n_xn_yn_z - n_x n_{y \wedge z}) = \frac{(n_x + n_y)(n_x + n_z)}{n_x} - \frac{(n_x + n_y)(n_x + n_z)n_{y \wedge z}}{n_xn_yn_z}, \\
     \gamma &= \frac{(n_x + n_y)(n_x + n_z)n_{y \wedge z}}{n_xn_yn_z} = \frac{n_x n_{y \wedge z}}{n_y n_z} + \frac{ n_{y \wedge z}}{n_y} + \frac{ n_{y \wedge z}}{n_z} + \frac{ n_{y \wedge z}}{n_x}. 
\end{align*}

For the first part, we determine the total number of quadruplets where $i_1= i_2, j_1 \neq j_2$. Given $i_1 = i_2$, there are $n_x$ possibilities. For $j_1 \neq j_2$, the number of combinations depends on the relationship between $n_y$ and $n_z$. If $n_y > n_z$, then we can select one element from $n_z$ and another from $n_y - 1$, yielding $n_z(n_y - 1) = n_y n_z - n_z$. If $n_y < n_z$, the summation splits into two parts: $\sum_{j_1=1}^{n_y}\sum_{j_2 = 1}^{n_y}$ and $\sum_{j_1 =1}^{n_y}\sum_{j_2 = n_y+1}^{n_z}$, with a the total number of terms $n_y(n_y-1) + n_y(n_z-n_y) = n_y^2 - n_y + n_yn_z -n_y^2 = n_yn_z - n_y$. Thus, the final number of elements is $n_yn_z - n_{y \wedge z}$, where $n_{y \wedge z} = \min(n_y, n_z)$. The remaining parts of the calculation are straightforward.

Given that $\gamma$ divided by $N = n_x + n_y + n_z$ approaches zero, we now have:
\begin{align*}
    \frac{\beta}{N} &= \frac{(n_x + n_y)(n_x + n_z)}{n_x(n_x + n_y + n_z)} - \gamma/N = \left( 1 + \frac{n_y}{n_x}\right)\left(\frac{n_x}{N} + \frac{n_z}{N}\right) - \frac{\gamma}{N}  \rightarrow  \left(1 +  \frac{1}{\lambda_{x,y}}\right)(\lambda_x + \lambda_z) .
    % = \left(1 +  \frac{1}{\lambda_{x,z}}\right)(\lambda_x + \lambda_y).
\end{align*}

Let $\alpha_{x,y,z} = \left(1 +  \frac{1}{\lambda_{x,y}}\right)(\lambda_x + \lambda_z)$. Thus we obtain: 
\begin{align*}
    Cov \left\{\frac{1}{\sqrt{N}}R_{\textnormal{diff},t_1}^{(x,y)}, \frac{1}{\sqrt{N}}R_{\textnormal{diff},t_2}^{(x,z)}\right\} = \frac{1}{K^2}\sum_{k_1,k_2=1}^K \alpha_{x,y,z}c_{t_1k_1t_2k_2}^{(x,y,z)} 
    % \left\{ \alpha_{x,y,z}c_{t_1k_1t_2k_2}^{X,Y,Z} 
    % + \beta_{x,y,z}d_{t_1k_1t_2k_2}^{X,Y,Z} \right\} 
    + o(1).
\end{align*}

Asymptotically, we conclude:
$$
\Sigma^{(x,y,z)}_{t_1,t_2} = Cov\left(R_{\textnormal{diff},t_1}^{(x,y)}, R_{\textnormal{diff},t_2}^{(x,z)} \right) = \frac{1}{K^2} \sum_{k_1,k_2=1}^K \alpha_{x,y,x}c_{t_1k_1t_2k_2}^{(x,y,z)}.
$$

\end{proof}

\subsection{Proof of Theorem \ref{thm2}\label{proof2}}
\begin{proof}

From Theorem~\ref{thm1}, we have the joint distribution of $(\bm{R}_{\textnormal{diff}}^{(x,y)}, \bm{R}_{\textnormal{diff}}^{(x,z)})$. Let $J = (1,1,\cdot,1)^T$. We can express the differences in the average ranks as:
\begin{align}
\label{Req}
    \Bar{R}_{y \cdot \cdot \cdot}^{(x,y)} - \Bar{R}_{x \cdot \cdot \cdot}^{(x,y)} &= \frac{1}{T} \sum_{t=1}^T \left\{ R_{y\cdot t \cdot}^{(x,y)} - R_{x\cdot t \cdot}^{(x,y)} \right\} = \frac{1}{T}J^T \bm{R}_{\textnormal{diff}}^{(x,y)},\\
    \Bar{R}_{z \cdot \cdot \cdot}^{(x,z)} - \Bar{R}_{x \cdot \cdot \cdot}^{(x,z)} &= \frac{1}{T} \sum_{t=1}^T \left\{ R_{z\cdot t \cdot}^{(x,z)} - R_{x\cdot t \cdot}^{(x,z)} \right\} = \frac{1}{T}J^T \bm{R}_{\textnormal{diff}}^{(x,z)}.
\end{align}

The expression in \eqref{Req} are normally distributed with the following mean and variance. $$\frac{T}{\sqrt{N}}(\Bar{R}_{y \cdot \cdot \cdot}^{(x,y)} - \Bar{R}_{x \cdot \cdot \cdot}^{(x,y)}) \sim \mathcal{N}\left(\frac{\sqrt{N}}{2}J^T\pmb{\theta}^{(x,y)}, J^T\Sigma^{(x,y)} J\right),$$
and 
$$\frac{T}{\sqrt{N}}(\Bar{R}_{z \cdot \cdot \cdot}^{(\bm{x},\bm{z})} - \Bar{R}_{x \cdot \cdot \cdot}^{(x,z)}) \sim \mathcal{N}\left(\frac{\sqrt{N}}{2}J^T \pmb{\theta}^{(x,z)}, J^T\Sigma^{(x,z)} J\right).$$
The covariance between these two quantities is given by:
\begin{align*}
    Cov\left(\frac{1}{T}J^T \bm{R}_{\textnormal{diff}}^{(x,y)}, \frac{1}{T}J^T \bm{R}_{\textnormal{diff}}^{(x,z)} \right) = \frac{1}{T^2} J^T \Sigma^{(x,y,z)} J.
\end{align*}

The theorem then follows from the fact that if $X_1$ and $X_2$ are two normal random variables with correlation $\rho$, the density of $max(X_1, X_2)$ is given by:
\begin{align*}
    f(x) &= f_1(-x) + f_2(-x), \\
    f_1(x) &= \frac{1}{\sigma_1}\phi\left(\frac{x + \mu_1}{\sigma_1}\right) \times \Phi\left(\frac{\rho (x + \mu_1)}{\sigma_1 \sqrt{1-\rho^2}} - \frac{x + \mu_2}{\sigma_2 \sqrt{1-\rho^2}}\right), \\
    f_2(x) &=  \frac{1}{\sigma_2}\phi\left(\frac{x + \mu_2}{\sigma_2}\right) \times \Phi\left(\frac{\rho (x + \mu_2)}{\sigma_2 \sqrt{1-\rho^2}} - \frac{x + \mu_1}{\sigma_1 \sqrt{1-\rho^2}}\right). 
\end{align*}

In our case, $X_1 = \frac{\bar{R}^{(x,y)}_{y\cdot\cdot\cdot}-\bar{R}^{(x,y)}_{x\cdot\cdot\cdot}}{\sqrt{\widehat{var}(\bar{R}^{(x,y)}_{y\cdot\cdot\cdot}-\bar{R}^{(x,y)}_{x\cdot\cdot\cdot})}}$ and $X_2 = \frac{\bar{R}^{(x,z)}_{z\cdot\cdot\cdot}-\bar{R}^{(x,z)}_{x\cdot\cdot\cdot}}{\sqrt{\widehat{var}(\bar{R}^{(x,z)}_{z\cdot\cdot\cdot}-\bar{R}^{(x,z)}_{x\cdot\cdot\cdot})}}$.

\end{proof}

\subsection{Proof of Theorem \ref{thm3}\label{proof3}}
\begin{proof}

We provide a brief outline of the proof, which can be extended to other cases similarly. First, consider the expectation of the estimator $\hat{c}_{t_1k_1t_2k_2}^{(x,y)}$:
\begin{align*}
E \left(\hat{c}_{t_1k_1t_2k_2}^{(x,y)}\right) &= \frac{1}{n_x} \sum_{i=1}^{n_x} E \left\{ \left(\frac{1}{n_y} \sum_{j=1}^{n_y} I(y_{jt_1k_1} < x_{it_1k_1}) - \frac{1 - \hat{\theta}_{t_1k_1}^{(x,y)}}{2}\right) 
\right. \\
    &\quad \quad \quad \quad
    \left.
    \left(\frac{1}{n_y} \sum_{j=1}^{n_y} I(y_{jt_2k_2} < x_{it_2k_2}) - \frac{1 - \hat{\theta}_{t_2k_2}^{(x,y)}}{2}\right)  \right\} \\
    &= \frac{1}{n_x} \sum_{i=1}^{n_x} E \left\{ \left(\frac{1}{n_y} \sum_{j=1}^{n_y} I(y_{jt_1k_1} < x_{it_1k_1}) - \frac{1}{n_x n_y}\sum_{i=1}^{n_x}\sum_{j=1}^{n_y} I(y_{jtk} < x_{itk}) \right)
    \right. \\
    &\quad \quad \quad \quad
    \left. 
    \left(\frac{1}{n_y} \sum_{j=1}^{n_y} I(y_{jt_2k_2} < x_{it_2k_2}) - \frac{1}{n_x n_y}\sum_{i=1}^{n_x}\sum_{j=1}^{n_y} I(y_{jtk} < x_{itk}) \right)  \right\} \\
    &= \frac{1}{n_x} \sum_{i=1}^{n_x} E \left\{ \left(\frac{1}{n_y} \sum_{j=1}^{n_y} I(y_{jt_1k_1} < x_{it_1k_1}) - E \left(\frac{1}{n_y}\sum_{j=1}^{n_y} I(y_{jtk} < x_{itk}) \right) \right)
    \right. \\
    &\quad \quad \quad \quad
    \left. 
    \left(\frac{1}{n_y} \sum_{j=1}^{n_y} I(y_{jt_2k_2} < x_{it_2k_2}) - E\left(\frac{1}{n_y}\sum_{j=1}^{n_y} I(y_{jtk} < x_{itk}) \right)  \right)\right\}
    \\
    &
    \rightarrow 
    Cov\left(G_{t_1k_1}(X_{t_1k_1}), G_{t_2k_2}(X_{t_2k_2})\right)
    =c_{t_1k_1t_2k_2}^{(x,y)}
    \\
    \end{align*}

Similarly, for the estimator $\hat{c}_{t_1k_1t_2k_2}^{(x,y,z)}$, we have:
\begin{align*}    
    E \left( \hat{c}_{t_1k_1t_2k_2}^{(x,y,z)}\right) &= \frac{1}{n_x} \sum_{i=1}^{n_x} E \left\{ \left(\frac{1}{n_y} \sum_{j=1}^{n_y} I(y_{jt_1k_1} < x_{it_1k_1}) - \frac{1 - \hat{\theta}_{t_1k_1}^{(x,y)}}{2}\right) 
    \right. \\
    &\quad \quad \quad \quad
    \left.
    \left(\frac{1}{n_z} \sum_{j=1}^{n_z} I(z_{jt_2k_2} < x_{it_2k_2}) - \frac{1 - \hat{\theta}_{t_2k_2}^{(x,z)}}{2}\right)  \right\}. \\
    % &= \frac{1}{n_x} \sum_{i=1}^{n_x} E \left\{ \left(\frac{1}{n_y} \sum_{j=1}^{n_y} I(y_{jt_1k_1} < x_{it_1k_1}) - \frac{1}{n_x n_y}\sum_{i=1}^{n_x}\sum_{j=1}^{n_y} I(y_{jtk} < x_{itk}) \right) \right. \\
    % &\quad \quad \quad \quad
    % \left. \left(\frac{1}{n_z} \sum_{j=1}^{n_z} I(z_{jt_2k_2} < x_{it_2k_2}) - \frac{1}{n_x n_z}\sum_{i=1}^{n_x}\sum_{j=1}^{n_z} I(z_{jtk} < x_{itk}) \right)  \right\} \\
    &= \frac{1}{n_x} \sum_{i=1}^{n_x} E \left\{ \left(\frac{1}{n_y} \sum_{j=1}^{n_y} I(y_{jt_1k_1} < x_{it_1k_1}) - E \left(\frac{1}{n_y}\sum_{j=1}^{n_y} I(y_{jtk} < x_{itk}) \right) \right)
    \right. \\
    &\quad \quad \quad \quad
    \left. 
    \left(\frac{1}{n_z} \sum_{j=1}^{n_z} I(z_{jt_2k_2} < x_{it_2k_2}) - E\left(\frac{1}{n_z}\sum_{j=1}^{n_z} I(z_{jtk} < x_{itk}) \right) \right)  \right\} 
    \\
    &
    \rightarrow 
    Cov\left(G_{t_1k_1}(X_{t_1k_1}), H_{t_2k_2}(X_{t_2k_2})\right)
    =c_{t_1k_1t_2k_2}^{(x,y,z)}.
\end{align*}
\end{proof}

\end{document}

\begin{frontmatter}

\title{Supplement for ``A Rank-Based Test for Multi-Arm Clinical Trials with Multivariate Longitudinal Outcomes"}

\author[1]{Dhrubajyoti Ghosh\corref{mycorrespondingauthor}}
\author[1]{Sheng Luo}

\address[1]{Duke University}
% \address[2]{Duke University}

\cortext[mycorrespondingauthor]{Corresponding author. Email address: dg302@duke.edu \url{}}
\end{frontmatter}

\section{BAPI Study Protocols}
The mean and standard devaitions for the placebo curve as taken from the BAPI 301 trial is given in Table 

\begin{table}[htbp]
\begin{tabular}{llll} \hline
                      &           & \multicolumn{1}{c}{Mean}                          & \multicolumn{1}{c}{SD}                            \\ \hline
\multirow{3}{*}{ADAS} & Placebo   & (21.99, 22.62, 23.69, 24.32, 25.57, 26.64, 27.38) & (9.96, 11.04, 12.17, 13.18, 14.42, 14.42, 15.27)  \\
                      & Low Dose  & (22.51, 22.94, 23.75, 24.45, 26.18, 26.22, 27.80) & (9.72, 10.75, 11.39, 12.46, 13.64, 14.14, 14.78)  \\
                      & High Dose & (22.04, 22.53, 23.52, 24.53, 25.73, 26.37, 27.31) & (9.67, 11.06, 12.07, 12.56, 13.50, 14.20, 14.56)  \\ \hline
\multirow{3}{*}{DAD}  & Placebo   & (80.57, 78.98, 77.46, 75.08, 72.45, 71.08, 68.00) & (19.09, 20.24, 21.58, 22.20, 24.29, 24.30, 26.60) \\
                      & Low Dose  & (80.71, 79.52, 78.25, 75.27, 74.18, 71.72, 69.60) & (17.64, 18.84, 20.61, 21.42, 23.06, 24.65, 24.29) \\
                      & High Dose & (80.85, 79.51, 77.95, 76.84, 73.23, 73.76, 70.03) & (18.77, 19.64, 20.63, 22.88, 25.02, 24.31, 27.48) \\ \hline
\end{tabular}
\caption{Mean and Standard Deviation of BAPI 301 Trial}
\end{table}

\section{Power Analysis Results}

\begin{table}[!h]
\begin{tabular}{|llllllllll|} \hline
   \backslashbox{Z.d}{Y.d}       & (0,0) & (0,1) & (0,3) & (0,5) & (0,5.39) & (0,6) & (1,0) & (1,1) & (1,3) \\ \hline
(0,0)    & 0.052  & 0.056  & 0.132  & 0.267  & 0.289     & 0.372  & 0.090  & 0.112  & 0.233  \\
(0,1)    & 0.089  & 0.093  & 0.171  & 0.243  & 0.282     & 0.475  & 0.147  & 0.168  & 0.234  \\
(0,3)    & 0.121  & 0.082  & 0.222  & 0.256  & 0.283     & 0.422  & 0.237   & 0.200  & 0.275  \\
(0,5)    & 0.268  & 0.298  & 0.325  & 0.347  & 0.312     & 0.479  & 0.354  & 0.232  & 0.334  \\
(0,5.39) & 0.219  & 0.358  & 0.361  & 0.343  & 0.443     & 0.524  & 0.389   & 0.378  & 0.483  \\
(0,6)    & 0.312  & 0.385  & 0.397  & 0.478  & 0.555      & 0.581  & 0.235  & 0.464  & 0.398  \\
(1,0)    & 0.192  & 0.197  & 0.199  & 0.294  & 0.323     & 0.387  & 0.116  & 0.190  & 0.274  \\
(1,1)    & 0.173  & 0.217  & 0.285  & 0.338  & 0.413      & 0.314  & 0.243  & 0.269  & 0.398  \\
(1,3)    & 0.236  & 0.295  & 0.366  & 0.392  & 0.334     & 0.417  & 0.394  & 0.326  & 0.419 \\ \hline
\end{tabular}
\caption{Power for different choices of deviations. The columns indicate deviation for dose 1, (1,2) meaning 1 unit improvement in ADAS and 2 in DAD. Similarly, the rows are for deviation in dose 2.\label{tab:first}
}
\end{table}

\begin{table}[htbp]
\begin{tabular}{|l|lllllllll|} \hline
   \backslashbox{Z.d}{Y.d}       & (0,0) & (0,1) & (0,3) & (0,5) & (0,5.39) & (0,6) & (1,0) & (1,1) & (1,3) \\ \hline
(1,5)      & 0.38  & 0.46  & 0.5   & 0.57  & 0.56     & 0.56  & 0.44  & 0.38  & 0.55  \\
(1,5.39)   & 0.42  & 0.43  & 0.47  & 0.55  & 0.52     & 0.54  & 0.52  & 0.51  & 0.57  \\
(1,6)      & 0.43  & 0.67  & 0.57  & 0.54  & 0.67     & 0.62  & 0.53  & 0.57  & 0.62  \\
(1.5,0)    & 0.12  & 0.2   & 0.16  & 0.27  & 0.27     & 0.41  & 0.19  & 0.21  & 0.25  \\
(1.5,1)    & 0.28  & 0.2   & 0.17  & 0.31  & 0.36     & 0.42  & 0.28  & 0.19  & 0.35  \\
(1.5,3)    & 0.29  & 0.49  & 0.32  & 0.54  & 0.42     & 0.48  & 0.36  & 0.39  & 0.53  \\
(1.5,5)    & 0.45  & 0.59  & 0.47  & 0.59  & 0.59     & 0.6   & 0.51  & 0.59  & 0.58  \\
(1.5,5.39) & 0.57  & 0.57  & 0.57  & 0.54  & 0.65     & 0.7   & 0.59  & 0.47  & 0.67  \\
(1.5,6)    & 0.64  & 0.63  & 0.55  & 0.68  & 0.58     & 0.55  & 0.62  & 0.68  & 0.62  \\ \hline
\end{tabular}
\end{table}

\begin{table}[htbp]
\begin{tabular}{|l|lllllllll|} \hline
   \backslashbox{Z.d}{Y.d}       & (0,0) & (0,1) & (0,3) & (0,5) & (0,5.39) & (0,6) & (1,0) & (1,1) & (1,3) \\ \hline
(2.21,0)    & 0.32  & 0.3   & 0.3   & 0.44  & 0.35     & 0.42  & 0.27  & 0.29  & 0.38  \\
(2.21,1)    & 0.24  & 0.31  & 0.29  & 0.48  & 0.45     & 0.47  & 0.3   & 0.33  & 0.4   \\
(2.21,3)    & 0.52  & 0.38  & 0.5   & 0.57  & 0.52     & 0.55  & 0.47  & 0.52  & 0.53  \\
(2.21,5)    & 0.69  & 0.67  & 0.72  & 0.72  & 0.74     & 0.76  & 0.65  & 0.65  & 0.71  \\
(2.21,5.39) & 0.66  & 0.7   & 0.74  & 0.74  & 0.72     & 0.72  & 0.66  & 0.69  & 0.74  \\
(2.21,6)    & 0.77  & 0.77  & 0.79  & 0.71  & 0.73     & 0.81  & 0.78  & 0.75  & 0.8   \\
(3,0)       & 0.39  & 0.4   & 0.36  & 0.46  & 0.56     & 0.54  & 0.38  & 0.42  & 0.39  \\
(3,1)       & 0.46  & 0.49  & 0.49  & 0.52  & 0.58     & 0.43  & 0.52  & 0.42  & 0.56  \\
(3,3)       & 0.71  & 0.65  & 0.66  & 0.73  & 0.66     & 0.77  & 0.61  & 0.73  & 0.69 \\ \hline
\end{tabular}
\end{table}

\begin{table}[htbp]
\begin{tabular}{|l|lllllllll|} \hline
   \backslashbox{Z.d}{Y.d}       & (0,0) & (0,1) & (0,3) & (0,5) & (0,5.39) & (0,6) & (1,0) & (1,1) & (1,3) \\ \hline
(3,5)    & 0.69  & 0.74  & 0.8   & 0.77  & 0.81     & 0.85  & 0.79  & 0.85  & 0.77  \\
(3,5.39) & 0.83  & 0.84  & 0.85  & 0.89  & 0.78     & 0.76  & 0.88  & 0.8   & 0.76  \\
(3,6)    & 0.87  & 0.9   & 0.92  & 0.86  & 0.91     & 0.84  & 0.87  & 0.88  & 0.95  \\
(5,0)    & 0.73  & 0.76  & 0.79  & 0.7   & 0.78     & 0.76  & 0.73  & 0.75  & 0.81  \\
(5,1)    & 0.78  & 0.84  & 0.85  & 0.87  & 0.85     & 0.88  & 0.84  & 0.89  & 0.83  \\
(5,3)    & 0.9   & 0.92  & 0.95  & 0.91  & 0.97     & 0.93  & 0.89  & 0.9   & 0.93  \\
(5,5)    & 0.98  & 0.97  & 0.93  & 0.98  & 0.97     & 0.99  & 0.97  & 0.97  & 0.97  \\
(5,5.39) & 0.97  & 0.97  & 0.96  & 0.98  & 0.96     & 1     & 0.98  & 0.98  & 0.97  \\
(5,6)    & 0.99  & 0.96  & 1     & 0.99  & 0.97     & 0.97  & 0.97  & 0.99  & 1 \\ \hline
\end{tabular}
\end{table}

\begin{table}[htbp]
\begin{tabular}{|l|lllllllll|} \hline
   \backslashbox{Z.d}{Y.d}      & (1,5) & (1,5.39) & (1,6) & (1.5,0) & (1.5,1) & (1.5,3) & (1.5,5) & (1.5,5.39) & (1.5,6) \\ \hline
(0,0)    & 0.36  & 0.51     & 0.47  & 0.22    & 0.16    & 0.36    & 0.48    & 0.62       & 0.6     \\
(0,1)    & 0.4   & 0.4      & 0.51  & 0.11    & 0.2     & 0.33    & 0.56    & 0.5        & 0.58    \\
(0,3)    & 0.4   & 0.48     & 0.6   & 0.17    & 0.42    & 0.37    & 0.57    & 0.56       & 0.57    \\
(0,5)    & 0.51  & 0.46     & 0.58  & 0.19    & 0.24    & 0.45    & 0.57    & 0.63       & 0.59    \\
(0,5.39) & 0.55  & 0.56     & 0.5   & 0.3     & 0.44    & 0.42    & 0.53    & 0.61       & 0.64    \\
(0,6)    & 0.56  & 0.52     & 0.58  & 0.39    & 0.4     & 0.5     & 0.63    & 0.63       & 0.69    \\
(1,0)    & 0.48  & 0.48     & 0.55  & 0.22    & 0.24    & 0.36    & 0.52    & 0.51       & 0.55    \\
(1,1)    & 0.51  & 0.41     & 0.53  & 0.21    & 0.3     & 0.43    & 0.58    & 0.59       & 0.59    \\
(1,3)    & 0.54  & 0.56     & 0.59  & 0.36    & 0.44    & 0.53    & 0.66    & 0.67       & 0.78  \\ \hline
\end{tabular}
\end{table}

\begin{table}[htbp]
\begin{tabular}{|l|lllllllll|} \hline
   \backslashbox{Z.d}{Y.d}      & (1,5) & (1,5.39) & (1,6) & (1.5,0) & (1.5,1) & (1.5,3) & (1.5,5) & (1.5,5.39) & (1.5,6) \\ \hline
(1,5)      & 0.61  & 0.54     & 0.71  & 0.44    & 0.5     & 0.53    & 0.61    & 0.54       & 0.71    \\
(1,5.39)   & 0.57  & 0.66     & 0.73  & 0.47    & 0.46    & 0.56    & 0.7     & 0.67       & 0.78    \\
(1,6)      & 0.61  & 0.64     & 0.74  & 0.54    & 0.51    & 0.6     & 0.74    & 0.76       & 0.77    \\
(1.5,0)    & 0.48  & 0.4      & 0.67  & 0.24    & 0.24    & 0.42    & 0.36    & 0.56       & 0.67    \\
(1.5,1)    & 0.54  & 0.45     & 0.58  & 0.21    & 0.27    & 0.49    & 0.58    & 0.63       & 0.66    \\
(1.5,3)    & 0.5   & 0.56     & 0.65  & 0.47    & 0.41    & 0.51    & 0.57    & 0.62       & 0.57    \\
(1.5,5)    & 0.68  & 0.63     & 0.72  & 0.56    & 0.67    & 0.6     & 0.66    & 0.67       & 0.78    \\
(1.5,5.39) & 0.72  & 0.71     & 0.64  & 0.68    & 0.57    & 0.68    & 0.72    & 0.8        & 0.8     \\
(1.5,6)    & 0.73  & 0.79     & 0.79  & 0.61    & 0.67    & 0.7     & 0.74    & 0.81       & 0.79    \\ \hline
\end{tabular}
\end{table}

\begin{table}[htbp]
\begin{tabular}{|l|lllllllll|} \hline
   \backslashbox{Z.d}{Y.d}      & (1,5) & (1,5.39) & (1,6) & (1.5,0) & (1.5,1) & (1.5,3) & (1.5,5) & (1.5,5.39) & (1.5,6) \\ \hline
(2.21,0)    & 0.53  & 0.49     & 0.44  & 0.19    & 0.34    & 0.44    & 0.53    & 0.53       & 0.74    \\
(2.21,1)    & 0.47  & 0.58     & 0.64  & 0.36    & 0.36    & 0.45    & 0.55    & 0.62       & 0.57    \\
(2.21,3)    & 0.65  & 0.61     & 0.64  & 0.51    & 0.51    & 0.65    & 0.66    & 0.65       & 0.75    \\
(2.21,5) & 0.69  & 0.7      & 0.81  & 0.67    & 0.63    & 0.75    & 0.79    & 0.79       & 0.8     \\
(2.21,5.39)    & 0.77  & 0.8      & 0.85  & 0.74    & 0.75    & 0.75    & 0.8     & 0.84       & 0.86    \\
(2.21,6)       & 0.86  & 0.78     & 0.81  & 0.78    & 0.84    & 0.77    & 0.79    & 0.84       & 0.85    \\
(3,0)       & 0.64  & 0.54     & 0.71  & 0.39    & 0.47    & 0.61    & 0.6     & 0.72       & 0.7     \\
(3,1)       & 0.72  & 0.59     & 0.66  & 0.5     & 0.54    & 0.55    & 0.65    & 0.65       & 0.74    \\
(3,3)       & 0.73  & 0.82     & 0.76  & 0.64    & 0.57    & 0.69    & 0.78    & 0.81       & 0.75 \\
\hline
\end{tabular}
\end{table}

\begin{table}[htbp]
\begin{tabular}{|l|lllllllll|} \hline
   \backslashbox{Z.d}{Y.d}      & (1,5) & (1,5.39) & (1,6) & (1.5,0) & (1.5,1) & (1.5,3) & (1.5,5) & (1.5,5.39) & (1.5,6) \\ \hline
(3,5)    & 0.87  & 0.86     & 0.85  & 0.81    & 0.78    & 0.85    & 0.82    & 0.8        & 0.86    \\
(3,5.39) & 0.85  & 0.86     & 0.84  & 0.87    & 0.84    & 0.91    & 0.87    & 0.85       & 0.89    \\
(3,6)    & 0.87  & 0.89     & 0.88  & 0.87    & 0.87    & 0.88    & 0.91    & 0.93       & 0.89    \\
(5,0)    & 0.8   & 0.85     & 0.83  & 0.73    & 0.74    & 0.76    & 0.79    & 0.84       & 0.87    \\
(5,1)    & 0.87  & 0.83     & 0.93  & 0.88    & 0.84    & 0.88    & 0.93    & 0.9        & 0.92    \\
(5,3)    & 0.9   & 0.95     & 0.92  & 0.96    & 0.96    & 0.91    & 0.94    & 0.93       & 0.94    \\
(5,5)    & 0.98  & 0.98     & 0.97  & 0.95    & 0.97    & 0.95    & 0.95    & 0.99       & 0.97    \\
(5,5.39) & 0.98  & 0.93     & 0.94  & 0.99    & 0.98    & 1       & 0.99    & 0.97       & 0.97    \\
(5,6)    & 1     & 0.98     & 0.98  & 0.98    & 0.98    & 0.97    & 0.98    & 0.99       & 0.98   \\
\hline
\end{tabular}
\end{table}